\documentclass[aps,prx,reprint,twocolumn,showpacs,superscriptaddress]{revtex4-2}
\usepackage[
colorlinks=true,
pdfborder={0 0 0},
breaklinks,
allcolors=blue
]{hyperref}
\usepackage{graphicx}
\usepackage{xcolor}
\usepackage{amsmath}
\usepackage{newtxtext}
\usepackage{orcidlink}
\usepackage{upgreek}

\newcommand{\vect}{\boldsymbol}

\newcommand{\Eqref}[1]{Eq.~\eqref{#1}}

\newcommand{\figref}[1]{Fig.~\ref{#1}}
\newcommand{\Figref}[1]{Fig.~\ref{#1}}
\newcommand{\kbT}{k_\mathrm{B}T}

\newcommand{\phiS}{\phi_\mathrm{S}}

\newcommand{\lC}{l_{P^+}}

\newcommand{\phiC}{\phi_{P^+}}
\newcommand{\lA}{l_{P^-}}

\newcommand{\phiA}{\phi_{P^-}}

\newcommand{\phip}{\phi_\mathrm{e^+}}
\newcommand{\phie}{\phi_\mathrm{e^-}}

\newcommand{\lb}{\ell_\mathrm{B}}

\newcommand{\br}{\mathrm{r}}

\newcommand{\phiAbar}{\bar{\phi}_{P^-}}
\newcommand{\phiCbar}{\bar{\phi}_{P^+}}

\newcommand{\FFH}{F_\mathrm{FH}}
\newcommand{\Fint}{F_\mathrm{int}}
\newcommand{\Fel}{F_\mathrm{el}}

\newcommand{\phibar}{\bar{\phi}}

\newcommand{\zC}{z_{P^+}}
\newcommand{\zA}{z_{P^-}}
\newcommand{\Lbeta}{{L_\beta}}

\newcommand{\phipp}{\phi_{P^+}}
\newcommand{\phipm}{\phi_{P^-}}
\newcommand{\Nc}{{N_\mathrm{c}}}

\newcommand{\zpp}{z_{P^+}}
\newcommand{\zpm}{z_{P^-}}
\newcommand{\Vbeta}{{V_\beta}}

\newcommand{\sectionCustom}[1]{}
\newcommand{\subsectionCustom}[1]{}

\begin{document}
\title{Condensate Size Control by Net Charge}

\author{Chengjie Luo\,\orcidlink{0000-0001-8443-0742}}
\thanks{These two authors contributed equally}
\affiliation{Max Planck Institute for Dynamics and Self-Organization, Am Faßberg 17, 37077 Göttingen, Germany}

\author{Nathaniel Hess\,\orcidlink{0000-0001-8971-332X}}
\thanks{These two authors contributed equally}
\affiliation{Department of Chemical and Biological Engineering, Princeton University, Princeton, NJ 08544, USA}

\author{Dilimulati Aierken\,\orcidlink{0000-0003-1727-5759}}
\affiliation{Department of Chemical and Biological Engineering, Princeton University, Princeton, NJ 08544, USA}
\affiliation{Omenn--Darling Bioengineering Institute, Princeton University, Princeton, NJ 08544, USA}

\author{Yicheng Qiang\,\orcidlink{0000-0003-2053-079X}}
\affiliation{Max Planck Institute for Dynamics and Self-Organization, Am Faßberg 17, 37077 Göttingen, Germany}

\author{Jerelle A. Joseph\,\orcidlink{0000-0003-4525-180X}}
\thanks{To whom correspondence may be addressed. Email: \href{mailto:jerellejoseph@princeton.edu}{jerellejoseph@princeton.edu} and \href{mailto:david.zwicker@ds.mpg.de}{david.zwicker@ds.mpg.de}}
\affiliation{Department of Chemical and Biological Engineering, Princeton University, Princeton, NJ 08544, USA}
\affiliation{Omenn--Darling Bioengineering Institute, Princeton University, Princeton, NJ 08544, USA}

\author{David Zwicker\,\orcidlink{0000-0002-3909-3334}}
\thanks{To whom correspondence may be addressed. Email: \href{mailto:jerellejoseph@princeton.edu}{jerellejoseph@princeton.edu} and \href{mailto:david.zwicker@ds.mpg.de}{david.zwicker@ds.mpg.de}}
\affiliation{Max Planck Institute for Dynamics and Self-Organization, Am Faßberg 17, 37077 Göttingen, Germany}

\date{\today}
\begin{abstract}
Biomolecular condensates are complex droplets comprising diverse molecules that interact using various mechanisms. Condensation is often driven by short-ranged attraction, but net charges can also mediate long-ranged repulsion. Using molecular dynamics simulations and an equilibrium field theory, we show that such opposing interactions can suppress coarsening so that many droplets of equal size coexist at equilibrium. This size control depends strongly on the charge asymmetry between constituents, while the strength of the short-ranged attractions has a weak influence. Essentially, droplets expel ions, so they cannot screen electrostatics effectively, implying droplets acquire a net charge and cannot grow indefinitely. Our work reveals how electrostatic effects control droplet size, which is relevant for understanding biomolecular condensates and creating synthetic patterns in chemical engineering.
\end{abstract}
\maketitle

\sectionCustom{Introduction}

Biomolecular condensates are complex droplets that are key for numerous cellular functions~\cite{Banani2017}.
They typically comprise diverse biomolecules, including nucleic acids and proteins, and form by phase separation due to various short-ranged interactions. On microscopic length scales, 
typical interactions include $\uppi$--$\uppi$-stacking, cation--$\uppi$ interactions, hydrogen bonding, and hydrophobic interactions~\cite{Holehouse2025,Dignon2020}.   %
Moreover, many of the involved molecules are charged to different degrees~\cite{Dignon2020}, implying electrostatic interactions.
These electrostatic interactions can contribute to the short-ranged interactions leading to phase separation.
An example are complex coacervates, where oppositely-charged polyions attract each other and phase separate from a solvent~\cite{Sing2020, Zhou2018, pakSequenceDeterminantsIntracellular2016, Meyer2024, lyonsFunctionalPartitioningTranscriptional2023}.
Electrostatics also induce interactions between macroscopic regions with net charge, although
 such long-ranged interactions are typically screened by mobile salt ions that neutralize charges on macromolecules in dilute electrolytes like the cytosol~\cite{Debye1923}.
However, screening might be suppressed if phase separation affects the distribution of ions.
Taken together, short-ranged interactions are crucial for forming biomolecular condensates, but the influence of long-ranged electrostatic effects is unclear.

Previous theoretical work suggests that long-ranged electrostatic effects are relevant in phase separating systems.
For instance, a net charge of droplets affects their interfaces~\cite{Posey2024,Majee2023}, reduces coarsening~\cite{Chen2022b, chenChargeAsymmetrySuppresses2023, Li2022a}, and can potentially lead to fission~\cite{Deserno2001,Rayleigh1882}.
In particular, reduced surface tension~\cite{anEffectsChargeAsymmetry2024} and an energy barrier suppressing coalescence~\cite{chenChargeAsymmetrySuppresses2023} slow down coarsening.
Simulations also find that heterogeneous charges can lead to multiphasic coacervates~\cite{chenMultiphaseCoacervationPolyelectrolytes2023}, and while polyelectrolytes with more charges phase separate more strongly~\cite{neitzelPolyelectrolyteComplexCoacervation2021}, a large charge asymmetry can suppress phase separation completely~\cite{anEffectsChargeAsymmetry2024}.
These observations indicate that net charge fundamentally affects phase separation of macromolecules, %
but the precise mechanism, and the role played by ions, is elusive.

\sectionCustom{MD simulations show droplet size control}
\begin{figure}
  \begin{center}
    \includegraphics[width=\columnwidth]{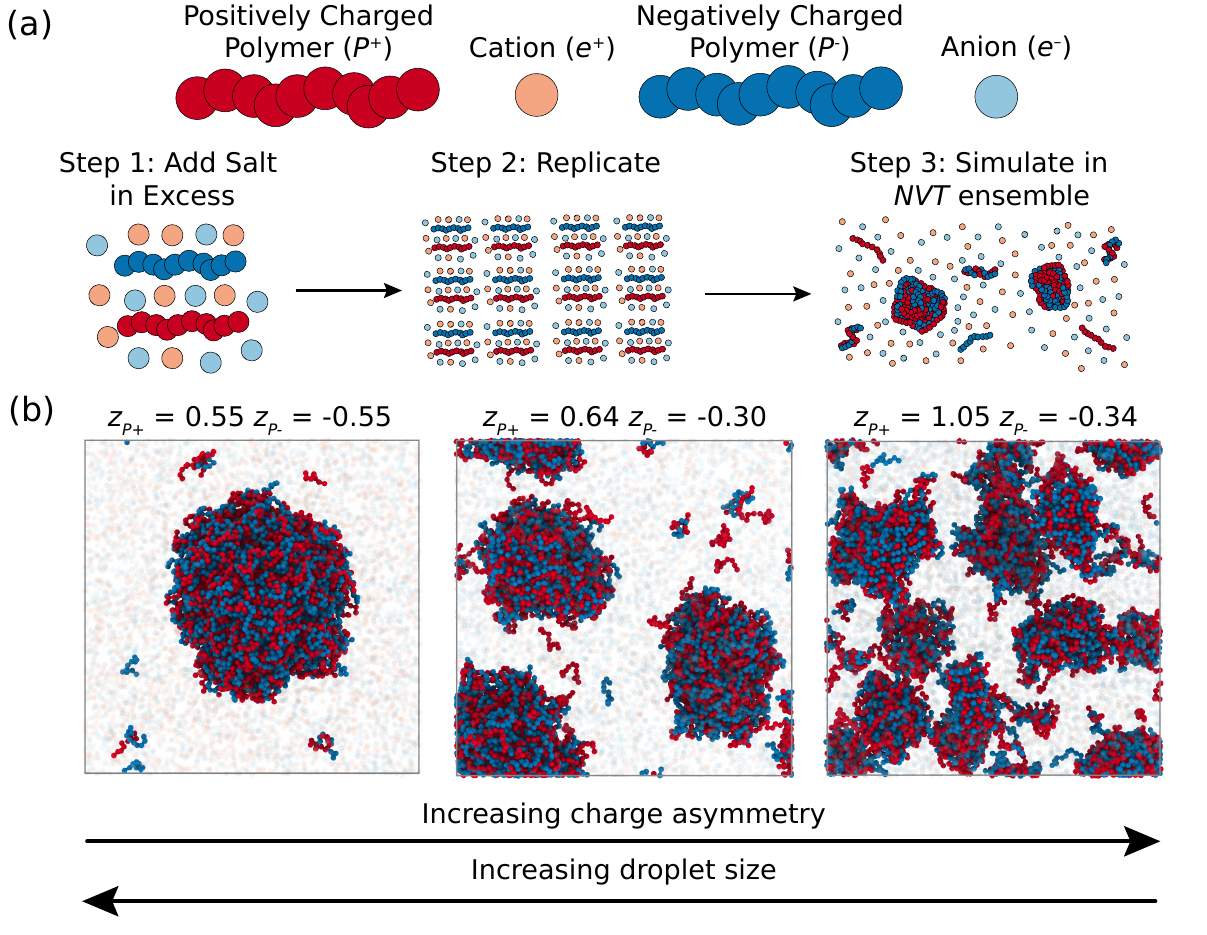}
    \caption{\textbf{MD simulations show smaller droplets for larger charge asymmetry.}
    (a) Schematic of simulations set-up, where a system of two oppositely charged polymers is neutralized with an excess of ions, replicated, and simulated in the $NVT$ ensemble (constant number of particles, volume, and temperature).
    (b) Snapshots from three simulations at various charges of the positively-charged ($\zC$, red) and negatively-charged polymers ($\zA$, blue) showing that larger charge asymmetry leads to smaller droplets.
    For clarity, ions are shown 99\% transparent.}
    \label{fig:Schematic}
  \end{center}
\end{figure}

To investigate phase separation of charged macromolecules in the presence of ions, we first performed molecular dynamics (MD) simulations of a system consisting of positively-charged polymers $P^+$, negatively-charged polymers $P^-$, and related ions, $e^+$ and $e^-$ in the \textit{NVT} ensemble with an implicit solvent modeled using a Langevin thermostat (\figref{fig:Schematic}a).
We induced phase separation using an attractive Lennard-Jones interaction between $P^+$ and $P^-$, while all other species exhibit excluded volume interactions.
For simplicity, we kept the length of the polymers fixed at ten times the size of an ion, and initialize the system such that $P^+$ and $P^-$ each occupy $10\%$ of the volume. 
Electrostatic interactions are given by Coulomb potentials between all species, which have charge numbers~$z_i$.
In particular, we varied the respective charge numbers $\zC>0$ and $\zA<0$ per monomer, whereas the ions have $z_{e^+}=1$ and $z_{e^-}=-1$.
To have comparable electrostatic effects, we added as many positive ions $e^+$ as there are charges on the negatively-charged polymers $P^-$, and vice versa for $e^-$.
Consequently, the overall system is neutral, and excess ions play the role of salt, which could in principle screen electrostatic effects between polymers.
Additional simulation details are described in the Supporting Text~\cite{supporting}.
The snapshots of equilibrated MD simulations shown in \Figref{fig:Schematic}b demonstrate that polymers form dense droplets surrounded by a dilute phase. %
Importantly, droplets are smaller for larger charge asymmetry $\Delta z = \zC + \zA$, indicating that charges can control droplet size and multiple droplets coexist stably.

\sectionCustom{Mean field theory reveals impact of charge asymmetry}

To reveal the physical mechanism of the observed droplet size regulation, we next describe the system using a continuous field theory.
Here, the state of the incompressible, isothermal mixture is captured by the volume fraction fields $\phiC(\vect r)$, $\phiA(\vect r)$, $\phip(\vect r)$, and $\phie(\vect r)$ of the charged species.
The remaining fraction, $\phiS=1-\sum_i \phi_i$ (here and below sums are over the four species $P^+$, $P^-$, $e^+$, and $e^-$), is filled by an inert solvent~$S$.
The system's equilibrium is governed by the minimum of the total free energy $F=\FFH + \Fint + \Fel$, where the terms respectively capture local, interfacial, and long-ranged electrostatic interactions akin to Ref.~\cite{Majee2023}.
We approximate local interactions using the Flory--Huggins model~\cite{Huggins1941,Flory1942},
\begin{equation}
    F_\mathrm{FH} =\frac{\kbT}{\nu}\!\int \bigg[
        \chi\phiA\phiC + 
        \phiS\ln\phiS + \sum_i\frac{\phi_i}{l_i} \ln\phi_i
    \biggl] \mathrm{d}V
 \;,
\end{equation}
where integrals are over the system of volume~$V$, $\kbT$ is the relevant thermal energy scale, and $\nu$ denotes the molecular volume, where solvent molecules, ions, and monomers of the polymers have the same volume for simplicity.
In the integrand, the first term proportional to the Flory parameter~$\chi<0$ captures the short-ranged attraction between $P^+$ and $P^-$, whereas the other terms capture translational entropies.
To mimic the MD simulations, we chose the respective chain lengths $l_i$ as $l_{e^+}=l_{e^-}=1$ and $\lC=\lA=10$.
The short-ranged interaction also implies an interfacial energy, which we describe by a simple gradient term~\cite{Cahn1958} 
\begin{equation}
    F_\mathrm{int} = \frac{\kappa\kbT}{2\nu} \!\int \sum_i |\nabla\phi_i|^2 \mathrm{d}V
    \;,
\end{equation}
which limits the width of interfaces between coexisting phases to roughly $\sqrt{\kappa}$ in strongly interacting systems~\cite{Weber2019}. %
Finally, long-ranged electrostatic effects are captured by the electrostatic potential~$\psi$, which is governed by the free energy
\begin{equation}
    F_\mathrm{el}=\int \biggl[-\frac{\varepsilon}{8\pi}|\nabla\psi|^2 + \frac{e\psi}{\nu}\sum_i z_i\phi_i\biggr]\mathrm{d}V
    \;,
\end{equation}
where $\varepsilon$ is the dielectric constant, assumed to be constant in space.
Taken together, the equilibrium state thus depends on the local attraction ($\chi$), the interfacial penalty ($\kappa$), and the charge numbers ($\zC$ and $\zA$).

\subsectionCustom{Mean field theory predicts patterned phase and macrophase separation. }

\begin{figure}
  \begin{center}
    \includegraphics[width=\columnwidth]{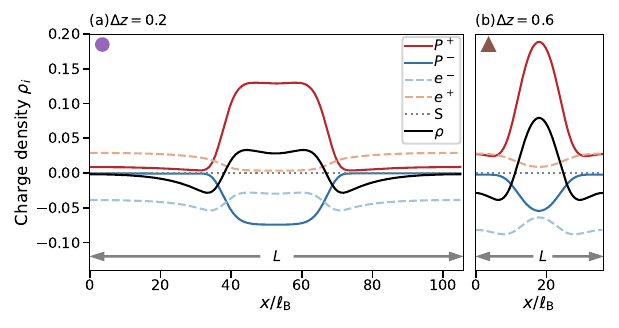}
    \caption{\textbf{Field theory predicts droplet-like periodic patterns.}
    Periodic profiles of charge density $\rho_i=z_i\phi_i$ for all species $i$ and total charge density $\rho=\sum_i \rho_i$ as a function of position~$x$ for weak (left) and strong (right)  charge asymmetry $\Delta z=\zC + \zA$.
    Parameters are $\zA=-0.2$, $\lC=\lA=10$, $l_{e^+}=l_{e^-}=1$, $\phiCbar=\phiAbar=0.1$, $\kappa=10\, \lb^2$, and $\chi=-5$, with Bjerrum length $\lb=e^2/(\varepsilon\kbT)$.
    }
    \label{fig:charge_profile_MF}
  \end{center}
\end{figure}

As in the MD simulations, we impose constant average fractions $\bar\phi_i = V^{-1} \int \phi_i \mathrm{d}V$ with $\phiCbar=\phiAbar=0.1$, $\phibar_{e^-}=\zC \phiCbar$, and $\phibar_{e^+}=-\zA \phiAbar$.
Based on the results shown in \figref{fig:Schematic}b, we expect that equilibrium states exhibit droplets of a well-defined size, which corresponds to periodic profiles in the field theory.
To describe such states, we consider a one-dimensional, periodic system of variable length $L$, and minimize $F$ by varying $L$, $\psi(x)$, and $\phi_i(x)$.
Here, we allow for coexisting phases with different periods and we impose mass conservation, incompressibility, and charge neutrality within each phase using Lagrange multipliers, akin to Ref.~\cite{Qiang2023} and described in the Supporting Text~\cite{supporting}.
The resulting equilibrium profiles shown in \figref{fig:charge_profile_MF} indicate that the field theory indeed predicts periodic patterns, where regions of large polymer density correspond to the droplets and the surrounding region enriched in ions represent the dilute phase.
Note that the ions do not neutralize the system everywhere (black line), which we confirmed in MD simulations (Fig. S1 in the Supporting Information~\cite{supporting}).
These net charges are reminiscent of the double layer structure of complex coacervates~\cite{Wang2019a} and indicate that electrostatics are crucial for droplet size control~\cite{Rubinstein2018}.
Indeed, we observe larger periods~$L$ for smaller charge asymmetry (\figref{fig:charge_profile_MF}), consistent with the larger droplets in our MD simulations (\figref{fig:Schematic}b).
There are also subtle differences between the two panels: For small charge asymmetry (\figref{fig:charge_profile_MF}a), the charge density~$\rho$ exhibits a dip inside the droplet and reaches neutrality ($\rho=0$) in the dilute phase.
Both effects are absent for larger charge asymmetry (\figref{fig:charge_profile_MF}b), suggesting that these two cases are qualitatively different.

\begin{figure*}
    \centering
    \includegraphics[width=\textwidth]{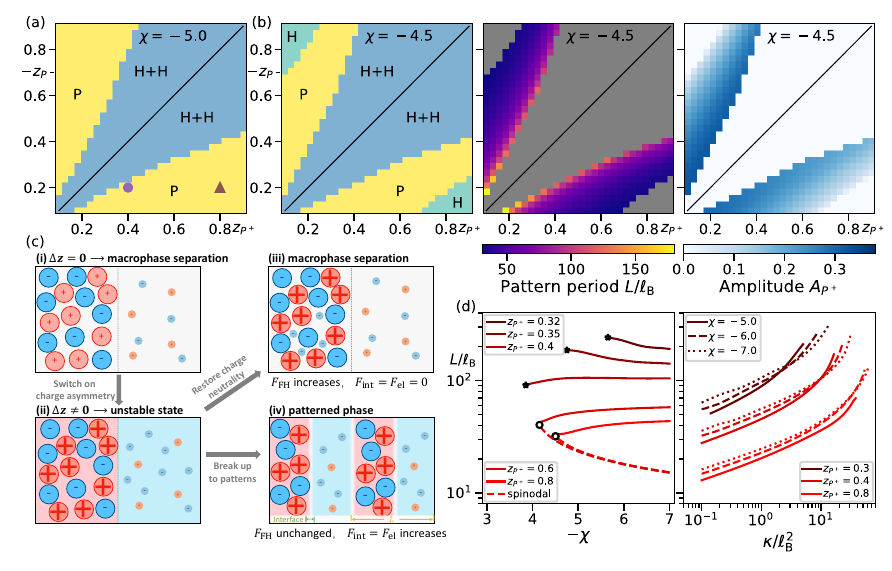}
    \caption{ 
    \textbf{Field theory predicts transitions to patterned phases.}
    (a) Phase diagram as a function of the charge numbers $\zC$ and $\zA$ of the polymers revealing parameter regions with the coexistence of two homogeneous phases (H+H) and patterned phases (P) for strong attraction ($\chi=-5$). The colored markers correspond to the two panels in \figref{fig:charge_profile_MF}.
    (b) Phase diagram for weak attraction ($\chi=-4.5$) including regions with a single homogeneous phase (H). 
    The remaining columns show respective periods~$L$ and amplitudes $A_{P^+} = \max(\phiC) - \min(\phiC)$. %
    (c) Schematic explanation of the origin of the patterned phase.
    The initial phase-separated charge-neutral state in (i) is destabilized due to net charges if charge asymmetry is enabled in (ii).
    An equilibrium state can either be reached by restoring charge neutrality in (iii) or by breaking the phases into smaller droplets in (iv).
    (d) Pattern length scale~$L$ as a function of interaction parameter $\chi$ (left, interfacial parameter $\kappa=10$) and $\kappa$ (right, various $\chi$) for various $\zC$ at $\zA=-0.2$.
    As $|\chi|$ decreases, the patterned phase transitions to macrophase separation (stars) or a homogeneous state (open disk).
    Dashed lines indicate the most unstable length scale predicted by the linear stability analysis in the Supporting Text~\cite{supporting}.
    (a--d) Additional parameters are given in \figref{fig:charge_profile_MF}.
     }
    \label{fig:types_MF} 
  \end{figure*}

We next analyze in detail how the periodic patterns depend on the charge numbers $\zC$ and $\zA$.
\figref{fig:types_MF}a shows that periodic patterns only appear in a restricted parameter region (yellow region P).
In particular, this patterned phase does not exist for symmetric mixtures (diagonal line, $\Delta z=0$).
Instead, the system exhibits macrophase separation (blue H+H region) if $\chi$ is sufficiently negative, and this region becomes larger for weaker interactions (\figref{fig:types_MF}b and Fig. S21~\cite{supporting}).
Moreover, the system stays homogeneous for very asymmetric mixtures (teal H region).
Collectively, these observations suggest  periodic patterns emerge for sufficiently strong attraction (large $-\chi$) and intermediate charge asymmetry $|\Delta z|$ between polymers.

\subsectionCustom{A simple argument explains the patterned phase}
We can understand the emergence of patterned phases starting from the charge-balanced state ($\Delta z=0$, \figref{fig:types_MF}c, subpanel i).
In this case, the polymers form a dense macrophase while small species ($e^+$, $e^-$, and $S$) accumulate in the corresponding dilute phase.
Here, we have $F= \FFH$ since $\Fel=0$ and the energy $\Fint$ of the single interface is negligible in a thermodynamically large system.
Switching on charge asymmetry ($\Delta z \neq 0$, subpanel ii) destabilizes this state because the macrophases acquire a net charge, so the electrostatic energy $\Fel$ diverges.
To minimize the total free energy~$F$, the state can change in two fundamentally different ways: it can either restore charge neutrality (subpanel iii) or form a patterned phase (subpanel iv).
In the first alternative, ions move into the dense phase (maintaining macrophase separation, subpanel iii), which restores charge neutrality and maintains $\Fel=\Fint=0$.
However, the ions decrease the concentration of polymers, thus reducing short-ranged contacts between polymers, and  increasing $\FFH$.
In the second alternative, the dense phase is broken up to form a pattern of finite length scale~$L$ (subpanel iv).
In the simplest case, the polymers maintain a high concentration, implying that $\FFH$ hardly changes.
However, the electrostatic energy $\Fel$ is now finite, and the breakup creates many interfaces, so both $\Fel$ and $\Fint$ are positive.
Taken together, both alternatives increase $F$ compared to the neutral initial case, but breakup into the patterned phase is favored for strong attraction (large $-\chi$).

The preceding argument also makes qualitative predictions for the pattern period~$L$ as a proxy for droplet size. %
Assuming that the density of polymers stays constant in the patterned phase, $\FFH$ is independent of $L$, so $L$ is instead governed by the minimum of $\Fint + \Fel$.
Generally, $\Fint$ decreases with increasing $L$ (since there are fewer interfaces per unit length), whereas $\Fel$ increases with $L$ (since opposite charges are separated further for larger $L$, i.e., exchanging the two central slabs in \figref{fig:types_MF}c(iv) costs electrostatic energy). 
These arguments predict that $L$ increases with larger $\kappa$ (increasing $\Fint$), whereas $L$ decreases with larger $|\Delta z|$ (increasing $\Fel$).
We present a more quantitative argument in the Supporting Text~\cite{supporting}, which shows that at the free energy minimum the two energies have opposing dependencies on $L$, $\mathrm{d}\Fint/\mathrm{d}L = -\mathrm{d}\Fel/\mathrm{d}L$, leading to a trade-off where $\Fint=\Fel$ at the minimum. %
We thus generally expect that $L$ increases with larger $\kappa$ and smaller $|\Delta z|$.

\subsectionCustom{Numerical minimizations confirm parameter dependencies}

We test our predictions by analyzing the patterned phases obtained from the field theory.
\figref{fig:types_MF}b shows that $L$ indeed decreases for larger $|\Delta z|$ (away from the symmetry line), although the patterns completely vanish for large $|\Delta z|$ where the homogeneous state exhibits the lowest free energy.
The right-most panel in \figref{fig:types_MF}b shows that the amplitude of $\phiC(x)$ decreases with increasing $|\Delta z|$, implying that polymers become less concentrated.
This is not captured by the simple argument above, suggesting that the system compromises the segregation of polymers (implying larger $\FFH$) for a lower electrostatic energy $\Fel$.
Indeed we find that $\Fel$ exhibits a non-monotonous dependency on $|\Delta z|$ (Supporting Figure~S11~\cite{supporting}), and the patterns cease when $\Fel$ vanishes.
Concomitantly, the amplitude approaches zero (\figref{fig:types_MF}b), suggesting that the transition to the homogeneous phase is continuous.

To also test the influence of the interaction strength~$\chi$ and the interfacial parameter~$\kappa$ on the pattern size $L$, we performed additional minimizations.
\figref{fig:types_MF}d confirms a weak influence of $\chi$, whereas $L$ increases significantly with $\kappa$.
Note that these length scales cannot be predicted from a simple stability analysis of the homogeneous state  (dashed line in \figref{fig:types_MF}d; details in the Supporting Text~\cite{supporting}), %
except at the transition of the patterned phase to the homogeneous state (open circles).
This observation, and the fact the scaling is consistent with mean-field theory (Supporting Figure S5~\cite{supporting}), provides additional evidence for a continuous phase transition. %
Taken together, the numerical solutions of the field theory confirm that $L$ (and thus the droplet size) increases with larger $\kappa$ and smaller $|\Delta z|$, while $\chi$ has a minor influence beyond driving droplet formation in the first place.

\sectionCustom{MD simulations corroborate parameter influence}

\begin{figure}
  \begin{center}
    \includegraphics[width=\columnwidth]{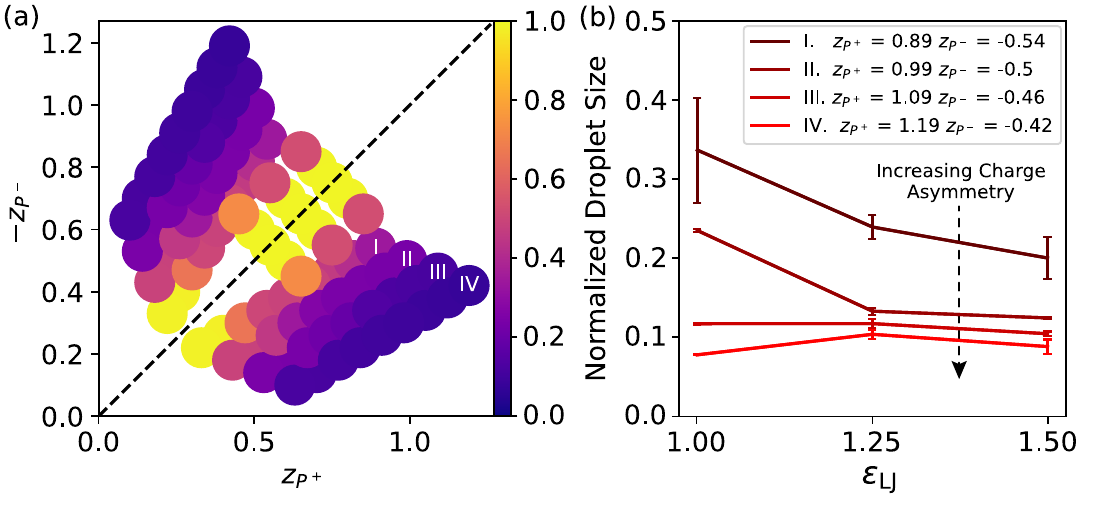}
    \caption{\textbf{MD simulations corroborate field theory.} (a) Simulations at various charge numbers ($\zC$, $\zA$) demonstrate that high charge asymmetries result in smaller average droplet sizes (normalized by total number of polymer chains in the system). (b) Normalized droplet size as a function of attraction strength $\epsilon_{\mathrm{LJ}}$. %
    Decreasing $\epsilon_{\mathrm{LJ}}$ mildly increases or decreases droplet size depending on charge asymmetry.}
    \label{fig:SimResults}
  \end{center}
\end{figure}

To test whether our predictions hold more generally, and in particular in 3D systems, we return to our MD simulations (details described in the Supporting Text~\cite{supporting}).
\figref{fig:SimResults}a shows that larger charge asymmetry $\Delta z$ indeed leads to smaller droplets.
We also observe system spanning droplets close to charge neutrality ($\Delta z=0$; yellow symbols), which are consistent with macrophase separation (region H+H in \figref{fig:types_MF}).
The detailed size distributions shown in Supporting Figure~S2~\cite{supporting} indicate this dependence unambiguously, although finite size effects are clearly visible for small charge asymmetries.
Moreover, \figref{fig:SimResults}b shows that varying the interaction strength~$\epsilon_\mathrm{LJ}$ between the polymers affects the droplet size only weakly, consistent with the weak influence of $\chi$ shown in \figref{fig:types_MF}d.
In both cases, at weak charge asymmetry droplet size slightly decreases for larger attraction, conversely to the behavior at strong charge asymmetry.
Overall, these data confirm the predictions from the field theory, suggesting that droplet size control by net charges is a general phenomenon.

\sectionCustom{Discussion}

One surprising feature of our model is the fact that salt, i.e., excess ions that are not necessary to neutralize the polymers, cannot screen the electrostatic interactions.
The concentration profiles obtained from MD simulations (Fig. S1~\cite{supporting}) and the field theory (Fig. \ref{fig:charge_profile_MF}) suggest that the interactions between polymers prevent ions from moving into the droplet to neutralize it, thus building up a net charge.
Consequently, the system cannot be described as a dilute electrolyte with mobile ions, e.g., using  Debye--Hückel theory~\cite{Debye1923}.
In particular, the intuitive argument that size-control would be limited by the Debye screening length is thus misleading.
Instead, ions need to be treated explicitly, which is likely a general feature of condensates involving complex, charged molecules~\cite{Smokers2024b,cheng2024mechanically} and similar systems~\cite{zhang2018salt,sadakane2009multilamellar,sadakane20112d}.

The described droplet size control essentially emerges from a trade-off between short-ranged attraction driving phase separation and long-ranged electrostatic repulsion if droplets accumulate net charges.
At weak charge asymmetry, the system exhibits macrophase separation since the attraction dominates.
Conversely, for strong charge asymmetry, the system will stay homogeneous to avoid accumulating net charges.
Between these two extremes, the patterned phases emerge as a compromise between phase separation and accumulating net charge.
Such patterned states also emerge in simpler phase separating system with Coulomb-like interactions~\cite{Ohta1986,Muratov2002,Tasios2017,onuki2016structure,Zhang2011a}.
In particular, when elastic effects mediate long-ranged repulsion, the patterned phase also exhibits a continuous transition to the homogeneous state~\cite{Qiang2023}.
This similarity suggests that droplet size can be generally controlled when the short-ranged attraction leading to phase separation is opposed by some long-ranged repulsion~\cite{Seul1995}, reminiscent of classical pattern formation~\cite{Cross2009}.
Another example for this mechanism are chemically active droplets, where long-ranged repulsion emerges from a reaction--diffusion system~\cite{Zwicker2015, Zwicker2017, Weber2019, Zwicker2022a}, which can in fact be interpreted using an electrostatic analogy~\cite{Zwicker2024,Ziethen2024,Liu1989}.

Here, we focused on the most basic mechanism of droplet size control by net charges.
For instance, we assumed that polymers have a fixed charge, whereas realistic molecules will exhibit charge regulation, so their behavior will depend on pH.
Molecules involved in typical biomolecular condensates are also much more complex: their charge distribution will be inhomogeneous, the nonspecific interactions will be heterogeneous and might depend on charge, and the polymer structure cannot be ignored, particularly for nucleic acids and intrinsically disordered protein regions~\cite{Holehouse2023}.
Such effects inform short-ranged interactions within biomolecular condensates~\cite{Banani2017,Dignon2020}, but the impact on long-ranged electrostatic is less understood.
Our work shows that such long-ranged  interactions can have profound consequences that cannot be explained by the simple Debye--Hückel theory. %
Therefore, it will be interesting to see how the described size control and ion distribution behaves in more complex systems as well as how these charge asymmetries affect condensates inside cells.
It is already tempting to speculate that charged droplets were implicated in the origin of life~\cite{Oparin1952,Brangwynne2012}, so that their size is maintained and the associated Rayleigh instability~\cite{Rayleigh1882,Deserno2001} could even explain division.

\begin{acknowledgments}
We thank Jan Kirschbaum for stimulating discussions, and we thank Nynke Hettema, Noah Ziethen and Athanassios Panagiotopoulos for helpful comments on our manuscript.
We gratefully acknowledge funding from the Max Planck Society and the European Union (ERC, EmulSim, 101044662), the Chan Zuckerberg Initiative DAF (an advised fund of Silicon Valley Community Foundation; grant 2023-332391 ), the National Institute of General Medical Sciences of the National Institutes of Health under Award Number R35GM155259, and departmental start-up funds via the Department of Chemical and Biological Engineering and the Omenn--Darling Bioengineering Institute at Princeton University. This research was partially supported by the National Science Foundation (NSF) through the Princeton University (PCCM) Materials Research Science and Engineering Center DMR-2011750. 
The simulations presented in this article were performed on computational resources managed and supported by Princeton Research Computing, a consortium of groups including the Princeton Institute for Computational Science and Engineering (PICSciE) and the Office of Information Technology's High Performance Computing Center.

\end{acknowledgments}

\appendix

\sectionCustom{Molecular Dynamics Simulations}
\label{sec:appendix_MD}

\bibliographystyle{apsrev4-2}
\bibliography{main}

\newpage

\onecolumngrid

\appendix

\renewcommand{\theequation}{S\arabic{equation}}\setcounter{equation}{0}

\renewcommand{\thefigure}{S\arabic{figure}}
  \setcounter{figure}{0}

  \section{Details of the molecular dynamics (MD) simulations}

  Here we present additional details on the MD simulations performed in this investigation.
  In particular, we provide additional details on the simulation procedure in subsection~\ref{sec:sim_procedure}.
  Additional figures pertaining to the data presented in the main text are presented in subsection~\ref{sec:sim_additional_figures}.
  In subsection~\ref{sec:sim_equilibration} we show data that demonstrates the simulations discussed in the main text are sampling equilibrium statistics.
  Further, we explore the impact of ion size on the mechanism of droplet size control by investigating a subset of simulations with reduced ion sizes in subsection~\ref{sec:sim_reduced_ion}.
  Finally, we validate that the damping constant of the Langevin thermostat does not alter simulation results if simulated in the overdamped limit (as opposed to the larger damping constant used in the simulations for the main text) in subsection~\ref{sec:sim_overdamped}.

  \subsection{Simulation procedure}
  \label{sec:sim_procedure}
  
  All simulations are performed in the MD software LAMMPS~\cite{thompsonLAMMPSFlexibleSimulation2022c}.
  MD simulations are run in the $NVT$ ensemble with neutral mixtures of charged polymers and ions in implicit solvent (Main Figure~1a).
  The equations of motion are integrated using the velocity-Verlet algorithm and the temperature is controlled using a Langevin thermostat (discussed in more detail below).
  All particles (i.e., monomers and ions) are modeled with a unit size $\sigma$ and a unit mass $m$. 
  Note that all simulation quantities reported in this SI and in the main text are in reduced Lennard--Jones units.
  
  The total energy of the simulation system is given by:
  \begin{equation}
      U = U_{\mathrm{el}}+U_{\mathrm{P}}+U_{\mathrm{EV}},
  \end{equation}
  where $U_{\mathrm{el}}$ describes the long-range electrostatic interactions between charged species (i.e., charged monomers on the polymers as well as ions), $U_{\mathrm{P}}$ describes the same chain bonded interactions and short-range heterotypic attraction between positively charged monomers $P^{+}$ and negatively charged monomers $P^{-}$, and $U_{\mathrm{EV}}$ describes the ion--ion, ion--polymer, and homotypic polymer--polymer excluded-volume interactions. 
  More specifically,  $U_{\mathrm{el}}$ is modeled using Coulomb's potential:
  \begin{equation}
       U_{\mathrm{el}} = \sum_{i,j}\frac{z_iz_j}{\varepsilon r_{ij}}.
  \end{equation}
  In this form, $z_i$ and $z_j$ describe the charge on species $i$ and $j$, $r_{ij}$ is the distance between the charged species, and the dielectric constant is $\varepsilon$. 
  Next, $U_{\mathrm{P}}$ models the combination of the bonded potential $ V_{\mathrm{bond}}$ and the Lennard-Jones potential $V_{\mathrm{LJ}}$ with cut-off and shifting:
  \begin{equation}
      U_{\mathrm{P}} = \sum_{i = 1}^{N} \sum_{j = 1}^{n-1} V_{\mathrm{bond}}(r_{j+1,j})+\sum_{i,j}\left(V_{\mathrm{LJ}}(r_{i,j})-V_{\mathrm{LJ}}(r_{\mathrm{c}})\right).
      \label{eq:sim U_P}
  \end{equation}
  Here, $N$ is total number of polymers in the system, $n$ is the polymer length.
  The bonded potential is modeled using a harmonic potential $V_{\mathrm{bond}}(r) = K(r-r_{\text{b}})^2$. 
  As for the Lennard-Jones potential we use:
  \begin{equation}
      V_{\mathrm{LJ}}(r) = 4\epsilon_{\mathrm{LJ}} \left[\left(\frac{\sigma}{r}\right)^{12}-\left(\frac{\sigma}{r}\right)^{6}\right],\, r\leq r_{\mathrm{c}}.
  \end{equation}
  Finally, for the excluded volume for all the particles (besides interactions between $P^+$ and $P^-$), we use the Weeks--Chandler--Andersen (WCA) potential
  \begin{equation}
      U_{\mathrm{EV}} = \sum_{i,j} \left[V_{\mathrm{LJ}}(r_{i,j})+\epsilon_{\mathrm{LJ}}\right], \,r_{i,j}\leq 2^{1/6} \sigma.
  \end{equation}
  
  In all simulations, we use a box size of $60 \times 60 \times 60 \: \sigma^{3}$ with our species volume fractions of $\phipp=\phipm=0.1$ and $\phi_{e^-}=\zpp \phipp$ and $\phi_{e^+}=-\zpm \phipm$.
  This corresponds to $N =512+512 =1024$ polymers where each polymer is composed of $n =10$ bonded monomers.
  We use a bond coefficient of $100.0\epsilon_{\mathrm{LJ}}$ and an equilibrium distance $r_{\mathrm{b}} = 1.0\sigma$. 
  For the Lennard-Jones interactions, we use a cut-off of $r_{\mathrm{c}} = 2.5\sigma$ and attraction strength of $\epsilon_{\mathrm{LJ}} = 1.0$. 
  A strength of $\epsilon = 1.0$ is also used for excluded volume  interactions of the WCA potential.
  The charge of polymers is varied by simulation. 
  The charge of ions is set to have a magnitude of $1.0$. 
  Electrostatic interactions are calculated using a PPPM grid~\cite{hockney2021computer} for interactions beyond $2.5\sigma$ with relative error in forces of less than $10^{-4}$. 
  Electrostatic interactions within $2.5\sigma$ are calculated directly. 
  The dielectric constant is set to $\varepsilon =80.0$. 
  
  Each simulation trajectory is first annealed at a temperature of $k_{\mathrm{B}}T/\epsilon_{\mathrm{LJ}} = 5.0$ for $10^4 \tau$ using a Langevin thermostat with a damping of $100.0 \tau$. 
  Thereafter, the temperature of each simulation is set to $k_{\mathrm{B}}T/\epsilon_{\mathrm{LJ}} = 1.0$ also using a Langevin thermostat with a damping of $100.0 \tau$. 
  Simulation trajectories are then run for $10^6\tau$. The last $5 \times10^{5}\tau$ is collected for data analysis, sampling a frame every $500\tau$. 
  We use a timestep of $0.005 \tau$ in the simulations.
  Additionally, all simulations in this study are performed in triplicate.
  Errors depicted in simulation figures correspond to the standard error between these simulation replicates during the sampling period of $5 \times10^{5}\tau$.

  In the analysis of MD simulations, clusters (droplets) are defined using a cutoff of $2.0\sigma$ between polymers. 
  The trajectory analysis software Ovito~\cite{ovito} is used to generate cluster data from simulation frames. 
  An in-house code is used to generate radial density profiles from an Ovito~\cite{ovito} cluster analysis pipeline.
  
  \subsection{Additional result figures}
  \label{sec:sim_additional_figures}
  
\begin{figure}
      \begin{center}
        \includegraphics[width=0.95\textwidth]{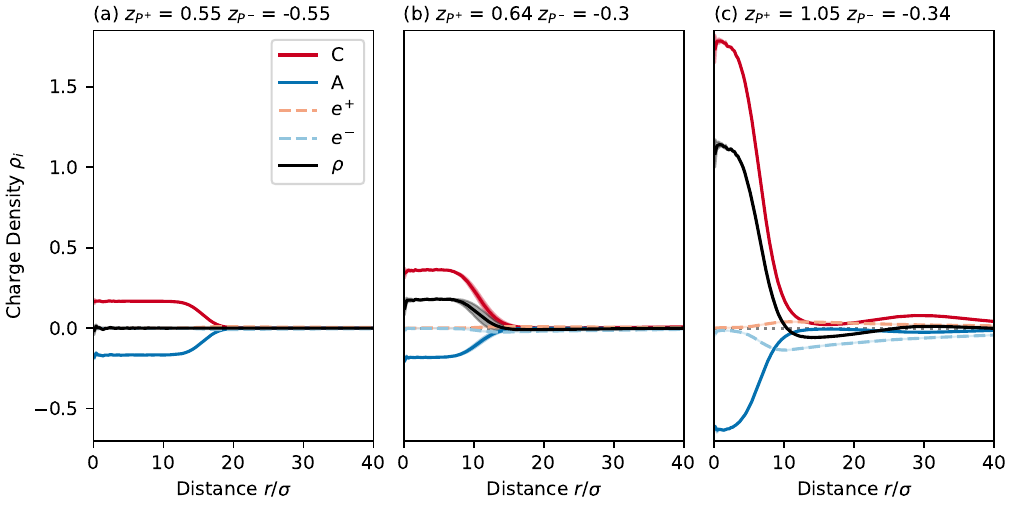}
        \caption{\textbf{Radial charge density profiles indicate charge asymmetry controls droplet size.} Radial charge density profiles for (a) $\zpp=0.55$, $\zpm=-0.55$ (b) $\zpp=0.64$, $\zpm=-0.3$ and (c) $\zpp=1.05$, $\zpm=-0.34$. Shaded regions indicate the standard error from three independent simulation replicates. The profiles indicate that increasing charge asymmetry decreases droplet size. They also demonstrate that ions are excluded from the dense phase. See Main Figure~1b for visualization of these droplets.}
        \label{fig:sim_rdps}
      \end{center}
    \end{figure}

\begin{figure}
      \begin{center}
        \includegraphics[width=0.8\textwidth]{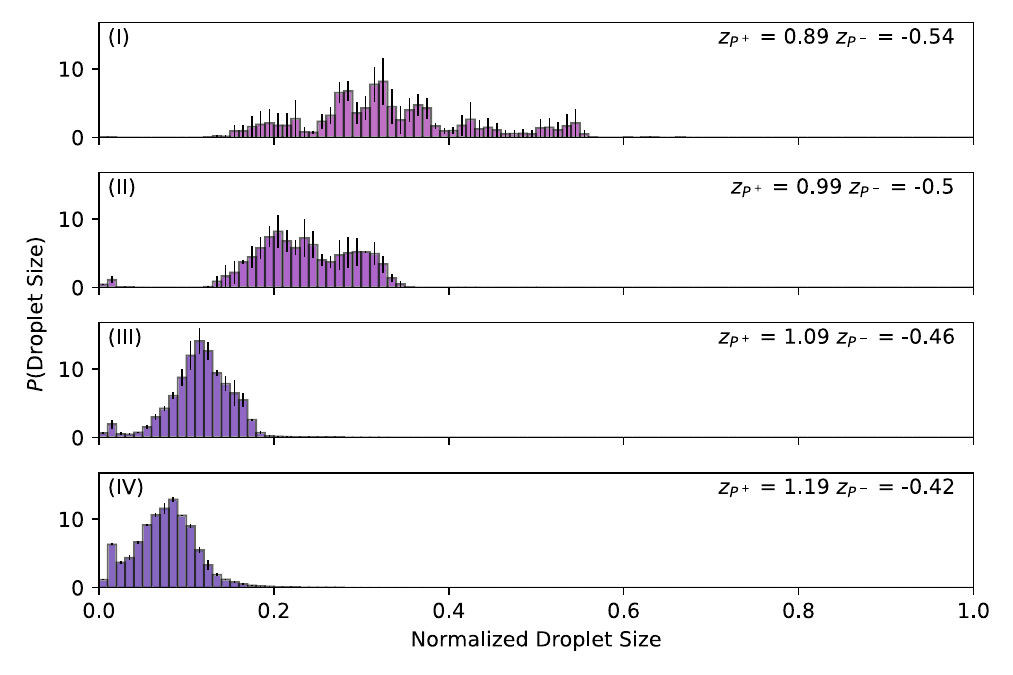}
        \caption{\textbf{Charge asymmetries create distributions of droplet sizes.} Four simulation systems, (I) $\zpp=0.89$, $\zpm=-0.54$ (II) $\zpp=0.99$, $\zpm=-0.5$ (III) $\zpp=1.09$, $\zpm=-0.46$ (IV) $\zpp=1.19$, $\zpm=-0.42$, corresponding to Main Figure 4a are plotted in histograms of normalized droplet sizes. Error bars indicate the standard error across histograms generated from three independent simulation replicates. At larger charge asymmetries (III and IV), a smaller average droplet size is favored. Finite-sized effects prevail at smaller charge asymmetries (I and II) indicated by a wide distribution of droplet sizes and high errors.}
        \label{fig:sim_hist}
      \end{center}
    \end{figure}

\begin{figure}
      \begin{center}
        \includegraphics[width=0.45\textwidth]{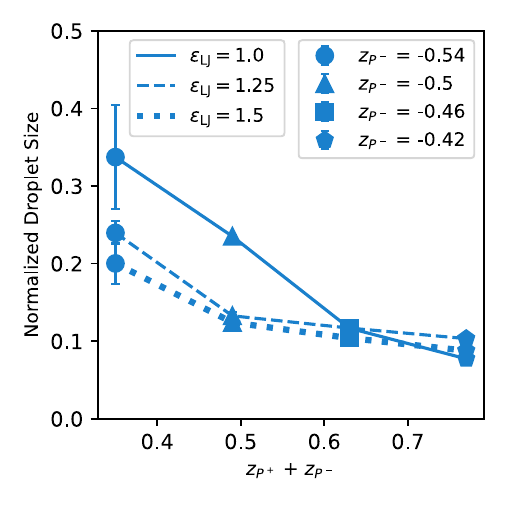}
        \caption{\textbf{Charge asymmetry's effect on droplet size is preserved at higher polymer attraction strengths.} The normalized droplet size for four charge pairings [($\zpp=0.89$, $\zpm=-0.54$), ($\zpp=0.99$, $\zpm=-0.5$), ($\zpp=1.09$, $\zpm=-0.46$), ($\zpp=1.19$, $\zpm=-0.42$) corresponding to I-IV in Main Figure~4a] at three different attraction strengths between oppositely charged polymers ($\epsilon_{\mathrm{LJ}} = 1.0,1.25,1.5$). As charge asymmetry is increased, the normalized droplet size in the simulations decreases regardless of the value of $\epsilon_{\mathrm{LJ}}$. Error bars represent the standard error in average normalized droplet size for three independent simulation replicates.}
        \label{fig:sim_charge_ep}
      \end{center}
    \end{figure}

\begin{figure}
      \begin{center}
        \includegraphics[width=0.45\textwidth]{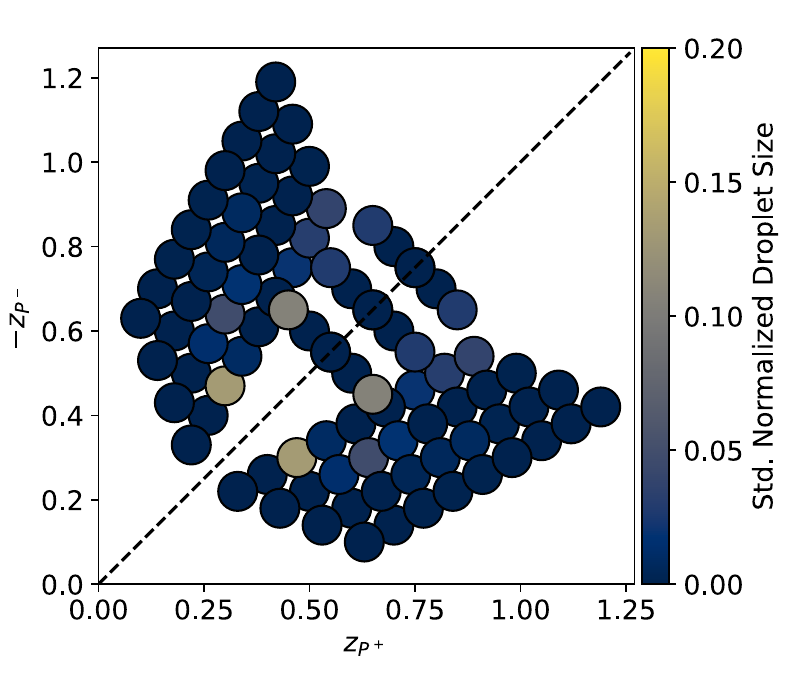}
        \caption{\textbf{Error on average simulation droplet sizes indicate finite-sized effects.}
        The standard error in the average droplet size (across three simulation replicates) for the charge pairings plotted in Main Figure~4a is shown. At intermediate charge asymmetries (charge asymmetries closer to the diagonal that still present a patterned phase), the presence of finite-sized effects is demonstrated by sustained droplets of different sizes (yellow/gray points).}
        \label{fig:sim_avg_error_sizes}
      \end{center}
    \end{figure}

\begin{figure}
      \begin{center}
        \includegraphics[width=0.45\textwidth]{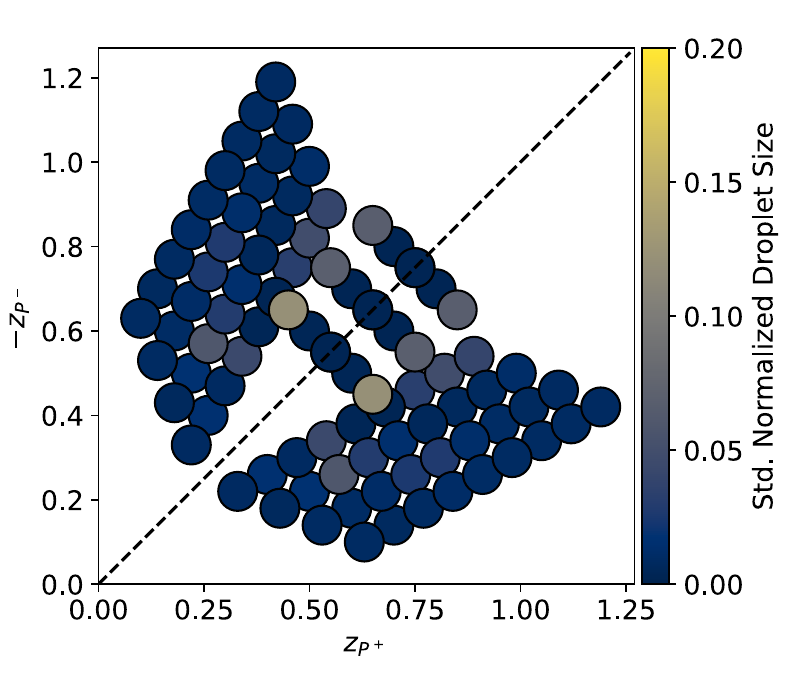}
        \caption{\textbf{Average standard deviation of droplet sizes within simulation replicates indicate finite-sized effects.}
        The standard deviation in the normalized droplet sizes plotted in Main Figure~4a averaged across three replicate simulations. Larger standard deviation (yellow/gray points) for moderate charge asymmetries are likely due to finite-sized effects resulting in size discrepancies among sustained droplets (see Fig.~\ref{fig:sim_hist}).}
        \label{fig:sim_std_sizes}
      \end{center}
    \end{figure}

  Additional figures are presented in this section corresponding to results from MD simulations discussed in the main text. 
  Figure~\ref{fig:sim_rdps} presents radial charge density profiles from MD simulations at ($\zpp=0.55$, $\zpm=-0.55$), ($\zpp=0.64$, $\zpm=-0.3$), and ($\zpp=1.05$, $\zpm=-0.34$). 
  See Main Figure~1b for a visualization of these droplets. 
  Figure~\ref{fig:sim_hist} shows normalized droplet size distributions corresponding to four charge pairings in Main Figure~4a. 
  Figure~\ref{fig:sim_charge_ep} analyzes the relationship between charge asymmetry and droplet size across three different values of $\epsilon_{\mathrm{LJ}}$. 
  Figure~\ref{fig:sim_avg_error_sizes} depicts the standard error in average droplet sizes across simulation replicates.
  Finally, Figure~\ref{fig:sim_std_sizes} quantifies the standard deviation in normalized droplet sizes within the simulations presented in Main Figure 4a.
  
  \clearpage
  
  \subsection{Demonstration of simulation equilibration}
  \label{sec:sim_equilibration}
  
  We consider a subset of the systems discussed in the main text to demonstrate our simulations have reached equilibrium.
  These simulations include ($\zpp=0.89$, $\zpm=-0.54$), ($\zpp=0.99$, $\zpm=-0.5$), ($\zpp=1.09$, $\zpm=-0.46$), and ($\zpp=1.19$, $\zpm=-0.42$) each of which was simulated with three independent replicates.
  To demonstrate equilibrium, we follow a three step process:
  
  \begin{enumerate}
      \item Begin with a simulation containing only 512 positively charged polymers and 512 negatively charged polymers at the same volume fraction discussed in subsection~\ref{sec:sim_procedure} above.
      Simulate this mixture in the $NpT$ ensemble with a Langevin thermostat ($T = 1.0$ and damping of $100.0\tau$) and a Berendsen barostat ($p = 5.0$ and damping of $100.0\tau$).
      This simulation is run for $2500\tau$.
      Turning off the Berendsen barostat, the simulation is progressed in the $NVT$ ensemble under the same Langevin thermostat for an additional $2500\tau$.
      In this simulation, we use Debye--H\"{u}ckel electrostatic interactions with a screening length of $5\sigma$ to ensure one large droplet is formed (i.e., $5\sigma$ is shorter than the predicted equilibrium droplet radii for all tested systems in the main text).
      This step produces one large droplet in a confined simulation box.
      \item Increase the box size to restore the proper volume fraction of polymers.
      Add the requisite number of ions to the simulation box to achieve the desired volume fractions (i.e., charge neutrality in excess salt).
      We then simulate in the $NVT$ ensemble under the same Langevin thermostat as above for $5000\tau$.
      For this step, we use Debye--H\"{u}ckel electostatic interactions with a screening length of $2\sigma$ to ensure that electrostatic repulsions are minimized and the single large droplet remains stable.
      \item Remove Debye--H\"{u}ckel screening. Run a $10^6\tau$ simulation, collecting the last $5 \times10^{5}\tau$ for analysis.
      This simulation is performed exactly as in subsection~\ref{sec:sim_procedure} but without the annealing step.
  \end{enumerate}
  
\begin{figure}
      \begin{center}
        \includegraphics[width=0.8\textwidth]{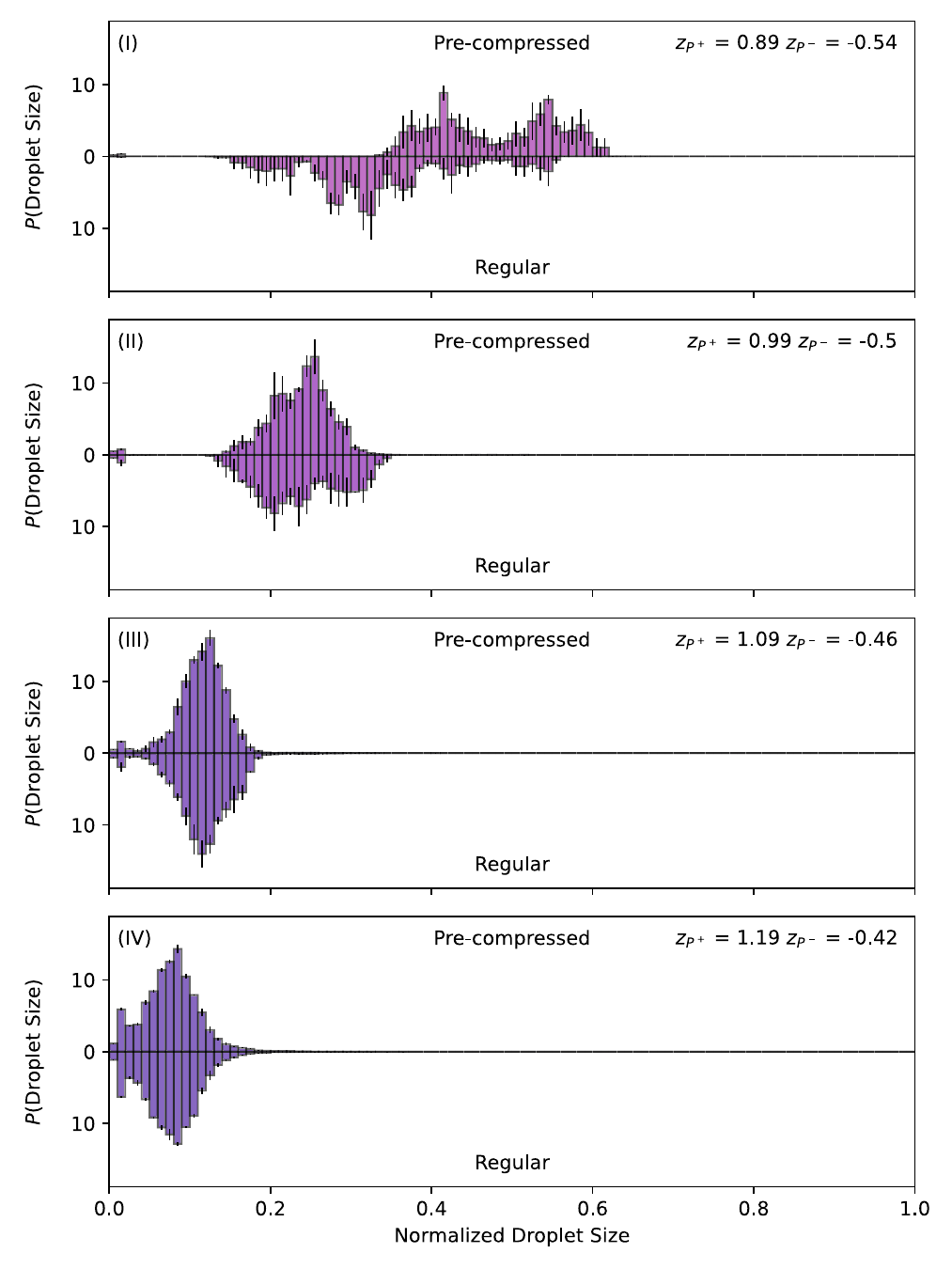}
        \caption{\textbf{Normalized droplet size distributions for pre-compressed versus regular simulations.} 
        The droplet size distributions for simulations that follow the `pre-compressed' (top) protocol (explained above) versus the `regular' (bottom) protocol (explained in subsection~\ref{sec:sim_procedure}) are presented.
        The distributions are very similar for both simulation protocols with the largest discrepancy appearing in the simulations for the charge pairing of ($\zpp=0.89$, $\zpm=-0.54$).
        The difference in the distributions for this simulation condition may be due to finite-sized effects creating long lasting droplets of various sizes that are difficult to quantitatively reproduce in two sets of simulation triplicates.}
        \label{fig:sim_hist_reg_eqm}
      \end{center}
  \end{figure}

\begin{figure}
      \begin{center}
        \includegraphics[width=0.8\textwidth]{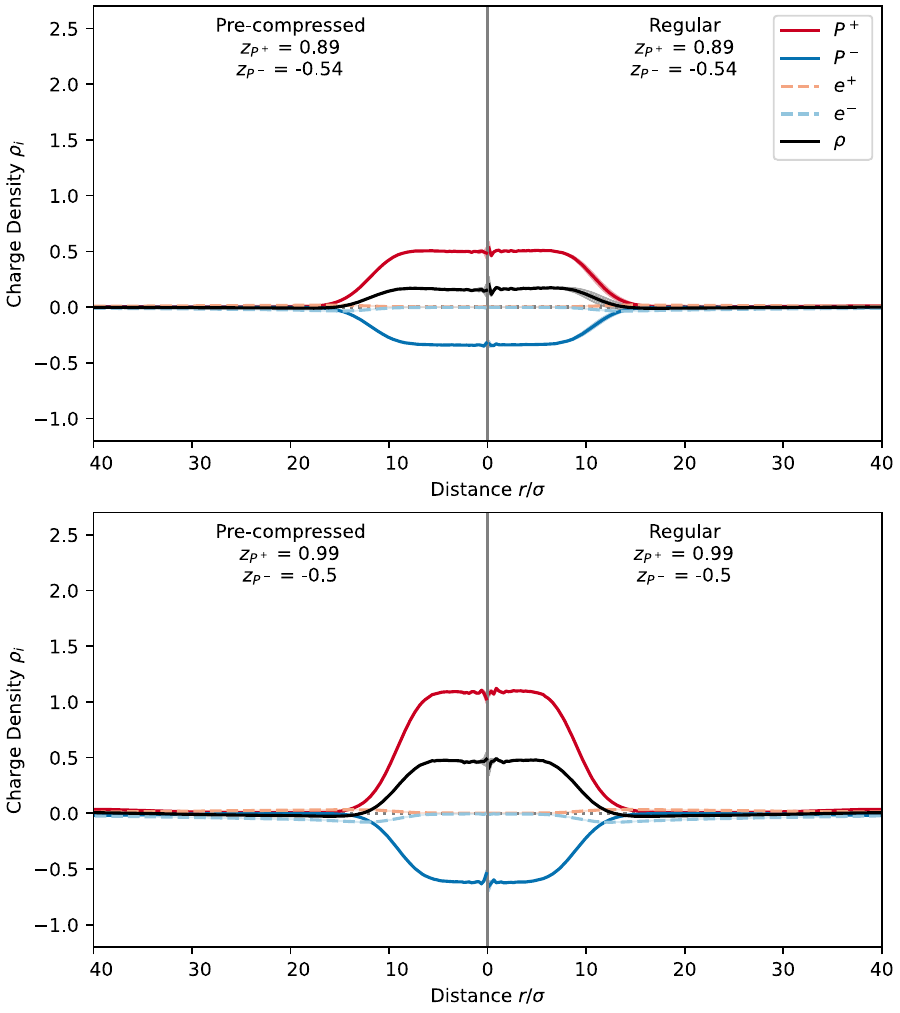}
        \caption{\textbf{Radial charge density profiles for pre-compressed versus regular simulations.} The radial density profiles are presented for the `pre-compressed' (left) versus `regular' (right) simulation protocols for the charge pairings of ($\zpp=1.09$, $\zpm=-0.46$) and ($\zpp=1.19$, $\zpm=-0.42$).
        The similarity between these distributions demonstrates that the regular simulations are sampling an equilibrated state.}
        \label{fig:sim_rdp_reg_eqm1}
      \end{center}
    \end{figure}

\begin{figure}
      \begin{center}
        \includegraphics[width=0.8\textwidth]{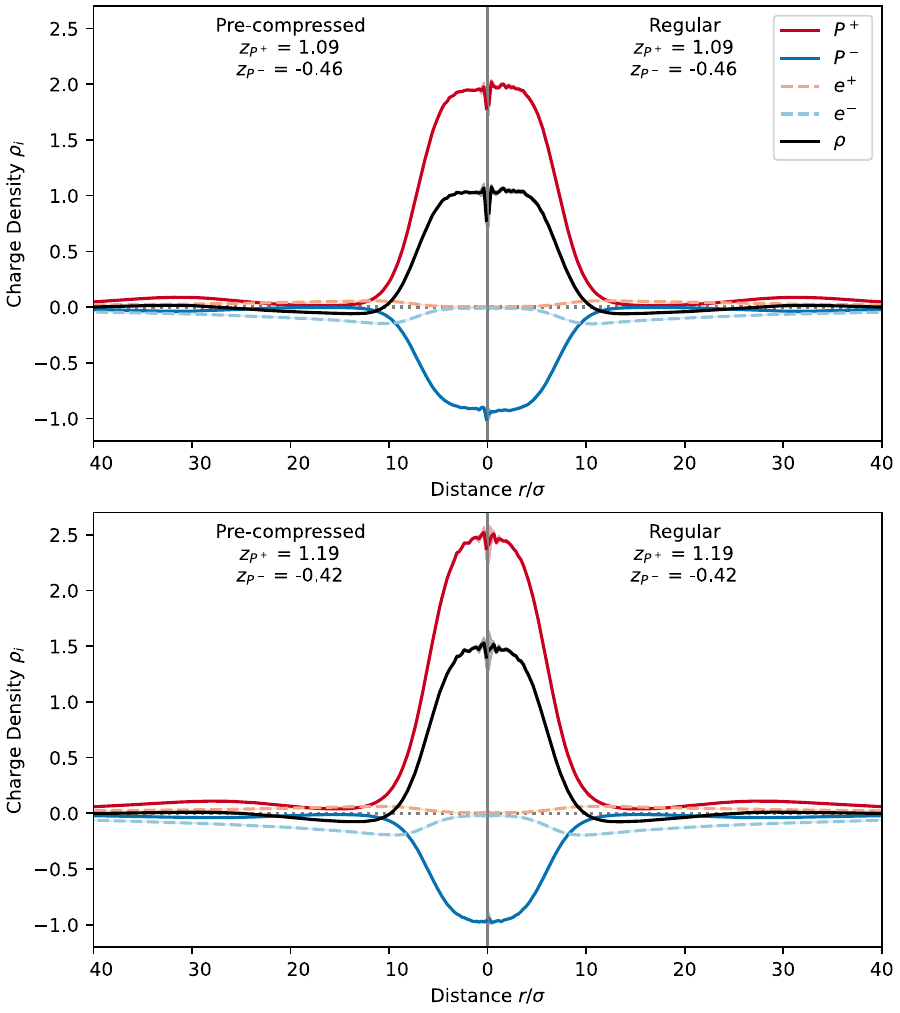}
        \caption{\textbf{Radial charge density profiles for pre-compressed versus regular simulations.} 
        The radial density profiles are presented for the `pre-compressed' (left) versus `regular' (right) simulation protocols for the charge pairings of ($\zpp=0.89$, $\zpm=-0.54$) and ($\zpp=0.99$, $\zpm=-0.5$).
        The similarity between these distributions demonstrates that the regular simulations are sampling an equilibrated state.}
        \label{fig:sim_rdp_reg_eqm2}
      \end{center}
    \end{figure}

  This process ensures that each simulation begins with one large droplet that contains all polymer chains, which is facilitated by the compression from the barostat and Debye--H\"{u}ckel screened electrostatic interactions with a short screening length.
  From this point, the simulations progress exactly as they are performed in subsection~\ref{sec:sim_procedure} but from this different initial state (one large droplet instead of thermal annealing). 
  Below we compare the results of these simulations to those presented in the main text and subsections \ref{sec:sim_procedure}-\ref{sec:sim_additional_figures} above.
  In particular, we compare the histograms of droplet sizes (Figure~\ref{fig:sim_hist_reg_eqm}) and the radial charge density profiles (Figures~\ref{fig:sim_rdp_reg_eqm1}, \ref{fig:sim_rdp_reg_eqm2}).
  The similarity in the results for simulations that follow the procedure outlined in this section versus those that follow the procedure outlined in subsection~\ref{sec:sim_procedure} provide strong evidence that we are sampling well-equilibrated systems.
  
  \clearpage
  
  \subsection{Simulations with reduced ion size}
  \label{sec:sim_reduced_ion}
  
  In this section we provide results that suggest the mechanism of size control via charge asymmetry is robust to ion size---at least with the parameters used in this investigation.
  To do this, we took a subset of simulations---($\zpp=0.89$, $\zpm=-0.54$), ($\zpp=0.99$, $\zpm=-0.5$), ($\zpp=1.09$, $\zpm=-0.46$), ($\zpp=1.19$, $\zpm=-0.42$)---and repeated our simulation procedure discussed in subsection~\ref{sec:sim_procedure} with a reduced ion size of $\sigma = 0.1$. 
  This reduces the volume of a single ion to be $1/1000\mathrm{th}$ the volume of a polymer monomer.
  The mass of the ions is still set to $m = 1$ for stability of the time integration (i.e., to preserve the same timestep as used previously).
  Additionally, in these simulations the interactions between polymers and ions are still modeled with the WCA potential, but the $\sigma$ values of cross-interactions are found using a geometric mixing rule.
  These repulsive interactions are also still modeled with an $\epsilon = 1.0$.
  Interactions between polymers are the same as discussed in subsection~\ref{sec:sim_procedure}.
  Finally, electrostatics are also treated identically to what was discussed in subsection~\ref{sec:sim_procedure}.
  
\begin{figure}
      \begin{center}
        \includegraphics[width=0.8\textwidth]{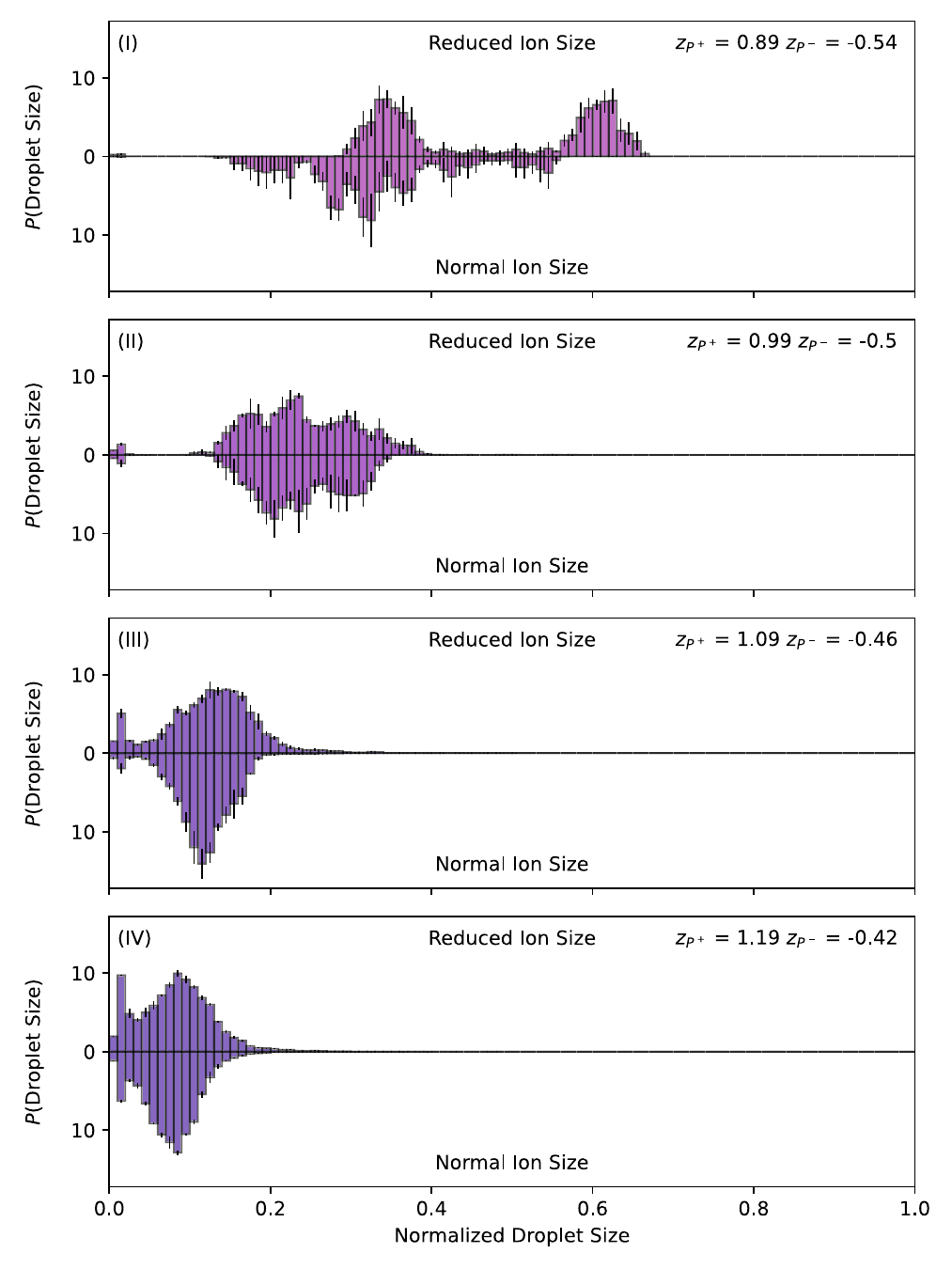}
        \caption{\textbf{Normalized droplet size distributions for reduced ion size versus normal simulations.} Droplet size histograms are presented for four charge pairings---($\zpp=0.89$, $\zpm=-0.54$), ($\zpp=0.99$, $\zpm=-0.5$), ($\zpp=1.09$, $\zpm=-0.46$), and ($\zpp=1.19$, $\zpm=-0.42$). For each charge pairing, the results are shown for three simulation replicates run with reduced ($\sigma = 0.1$) and normal ($\sigma = 1.0$) ion sizes. The results are qualitatively similar in both cases; even at a reduced ion size charge asymmetry can still control droplet sizes.}
        \label{fig:sim_hist_smallions}
      \end{center}
    \end{figure}

\begin{figure}
      \begin{center}
        \includegraphics[width=0.8\textwidth]{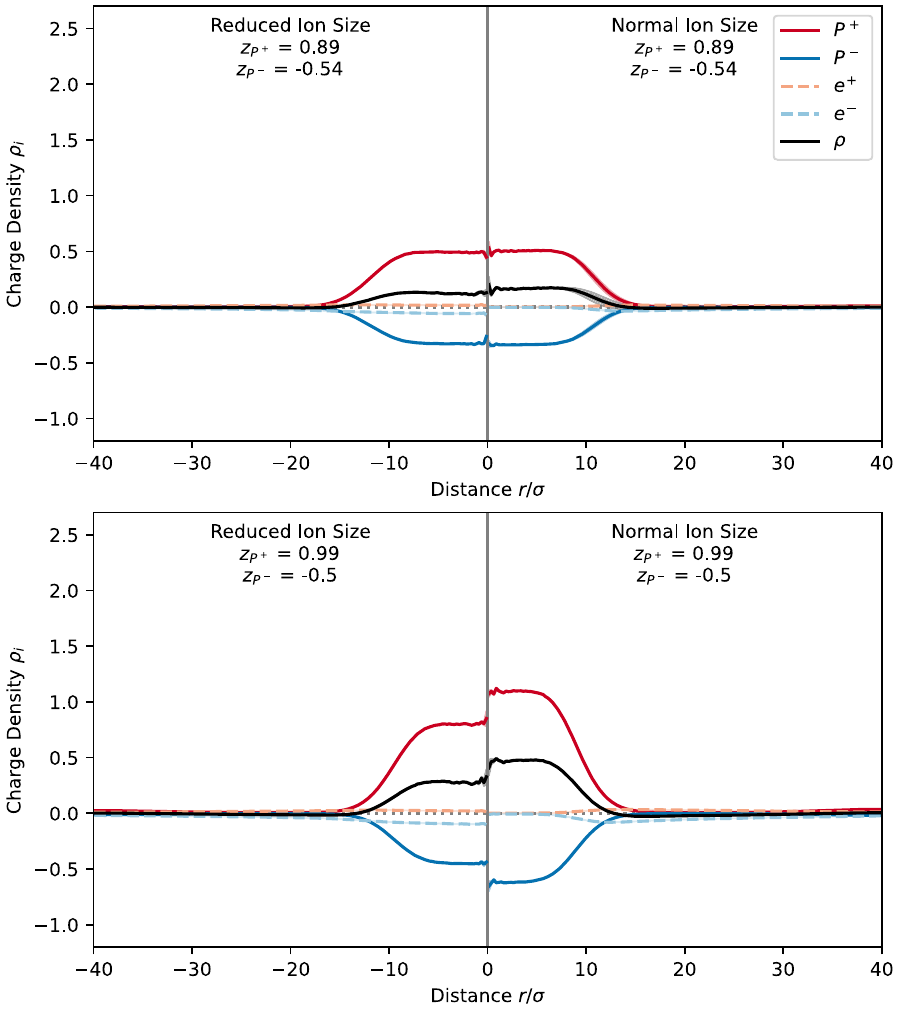}
        \caption{\textbf{Radial charge density profiles for reduced ion size versus normal simulations.} Radial density profiles are presented for the charge pairings ($\zpp=0.89$, $\zpm=-0.54$) and ($\zpp=0.99$, $\zpm=-0.5$) at both reduced ion size (left) and normal ion size (right) simulations. The results demonstrate that smaller ion sizes lead to more penetration of ions into the dense phase and a lower net charge density in the droplets. However, despite the increased concentration of ions in the dense phase compared to simulations with a normal ion size, the ions still cannot partition to the extent needed to neutralize droplet charge. Thus the mechanism of size control via producing a patterned phase still persists under these conditions.}
        \label{fig:sim_rdp_smallions1}
      \end{center}
    \end{figure}

\begin{figure}
      \begin{center}
        \includegraphics[width=0.8\textwidth]{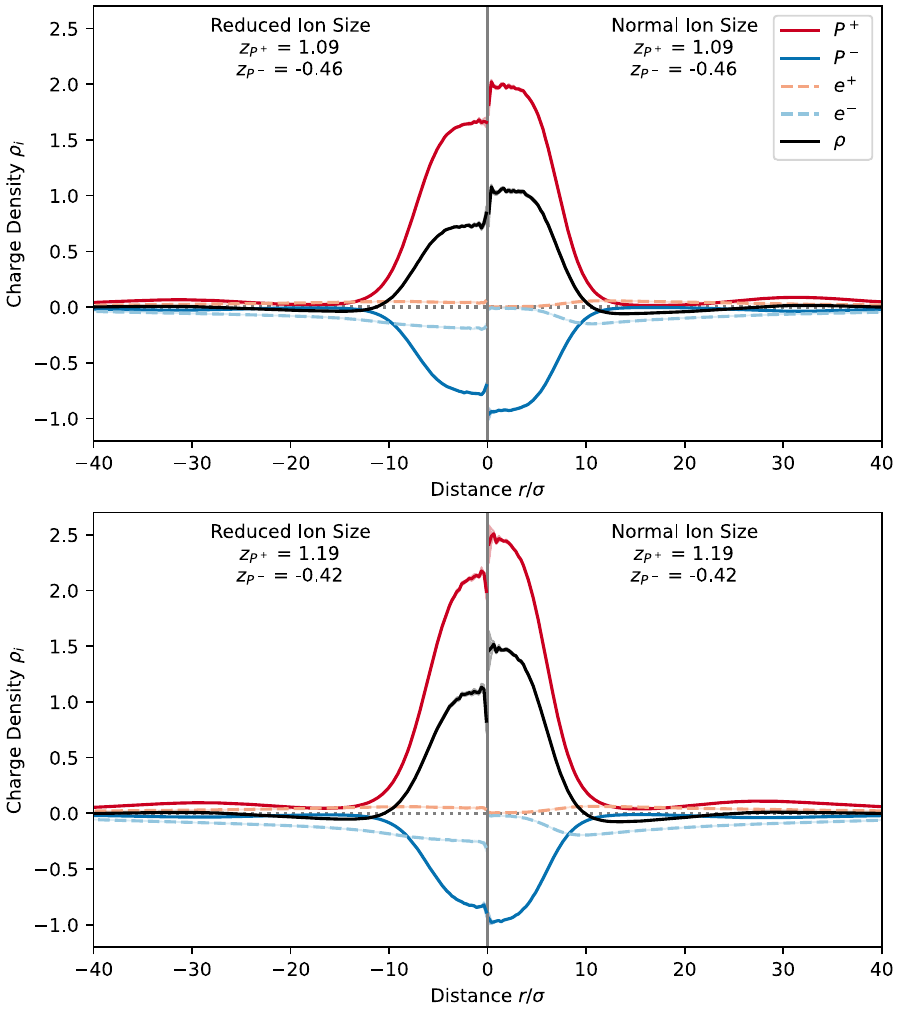}
        \caption{\textbf{Radial charge density profiles for reduced ion size versus normal simulations.} Radial density profiles are presented for the charge pairings ($\zpp=1.09$, $\zpm=-0.46$) and ($\zpp=1.19$, $\zpm=-0.42$) at both reduced ion size (left) and normal ion size (right) simulations. The results demonstrate that smaller ion sizes lead to more penetration of ions into the dense phase and a lower net charge density in the droplets. However, despite the increased concentration of ions in the dense phase compared to simulations with a normal ion size, the ions still cannot partition to the extent needed to neutralize droplet charge. Thus the mechanism of size control via producing a patterned phase still persists under these conditions.}
        \label{fig:sim_rdp_smallions2}
      \end{center}
    \end{figure}

  The simulations of reduced ion size were repeated in triplicate.
  The results presented below compare these simulations with reduced ion size to those performed under the regular simulation protocol outlined in subsection~\ref{sec:sim_procedure}.
  These results include a comparison of the normalized droplet size histograms (Figure~\ref{fig:sim_hist_smallions}) and comparisons of radial charge density profiles (Figures~\ref{fig:sim_rdp_smallions1}, \ref{fig:sim_rdp_smallions2}).
  The similarity between the droplet size histograms and the charge density profiles indicate that the mechanism of droplet size regulation may be robust to ion size, at least under the conditions used in this investigation.
  However, the most notable difference caused by the reduced ion size is the increased penetration of ions into the dense phase (Figures~\ref{fig:sim_rdp_smallions1}, \ref{fig:sim_rdp_smallions2}).
  This may indicate that as the discrepancy in polymer and ion sizes become quite large, ions will be able to invade the dense phase and neutralize the droplet thereby enabling full coarsening.
  Thus the exact details of any given system of interest (magnitude of charge asymmetry, ion size, polymer length and interaction strength, etc.) may be very important in determining whether electrostatic interactions can contribute to droplet size control.
  
\begin{figure}
      \begin{center}
        \includegraphics[width=0.8\textwidth]{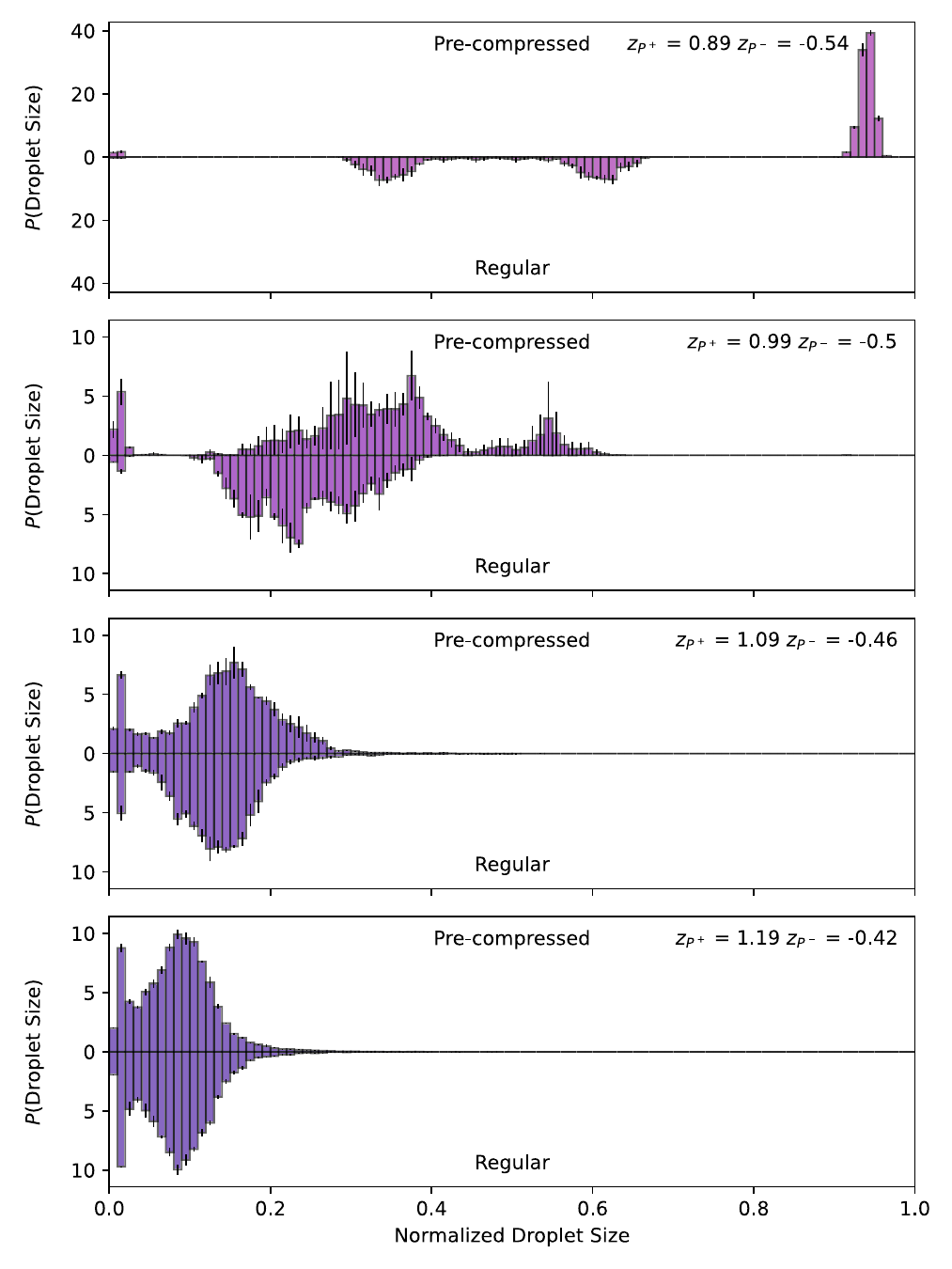}
        \caption{\textbf{Normalized droplet size distributions with reduced ion sizes for both pre-compressed and regular simulations.} Simulations with reduced ion size were performed using the pre-compressed (top) (subsection~\ref{sec:sim_equilibration}) and the regular (bottom) (subsection~\ref{sec:sim_procedure}) simulation procedures. Similarities between the droplet size distributions for the pre-compressed and regular simulations in the charge pairings ($\zpp=0.99$, $\zpm=-0.5$), ($\zpp=1.09$, $\zpm=-0.46$), and ($\zpp=1.19$, $\zpm=-0.42$) indicate that these simulations are likely sampling equilibrium statistics. For the ($\zpp=0.89$, $\zpm=-0.54$) charge pairing, discrepancy between the pre-compressed simulations---which shows one large droplet---and the regular simulations---which shows a distribution of smaller droplet sizes---indicates that for this charge pairing at least one of the two cases is not sampling equilibrium.}
        \label{fig:sim_hist_smallions_eqm}
      \end{center}
  \end{figure}

\begin{figure}
      \begin{center}
        \includegraphics[width=0.8\textwidth]{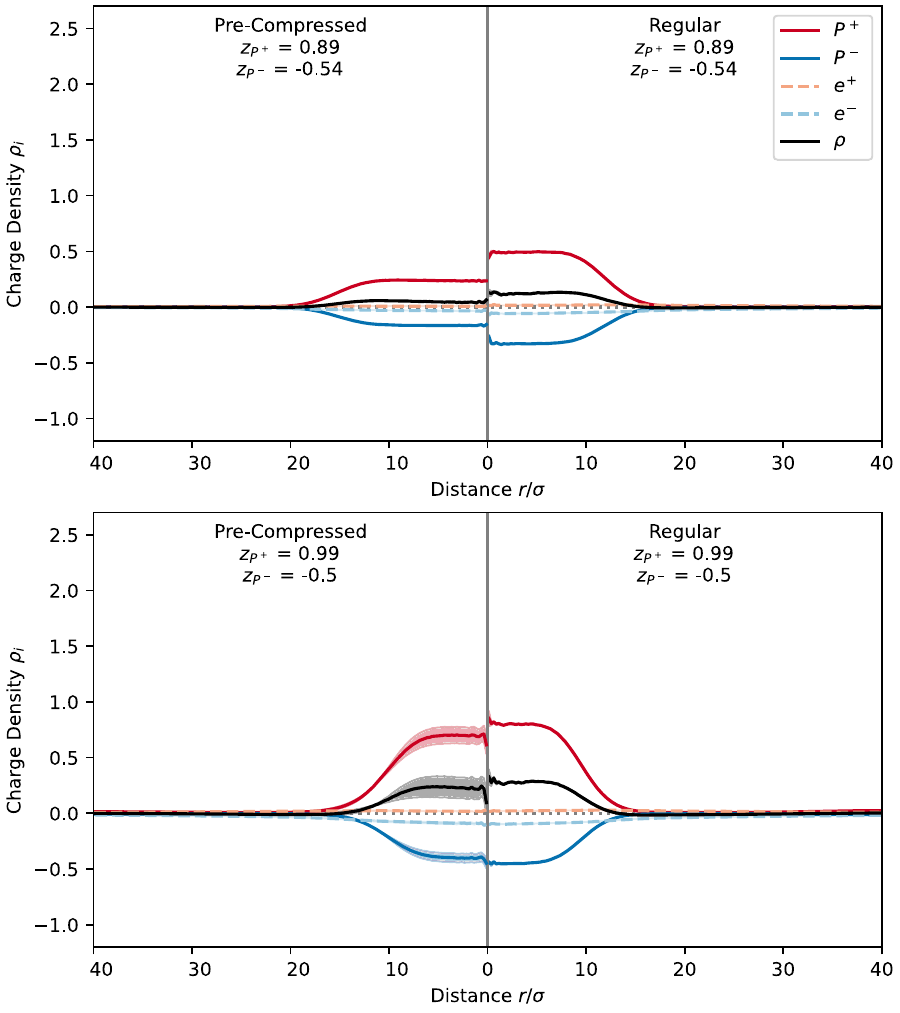}
        \caption{\textbf{Radial charge density profiles with reduced ion size for both pre-compressed and regular simulations.} Simulations with reduced ion size were performed using the pre-compressed (left) (subsection~\ref{sec:sim_equilibration}) and the regular (right) (subsection~\ref{sec:sim_procedure}) simulation procedures. Profiles are shown for the charge pairings of ($\zpp=0.89$, $\zpm=-0.54$) and ($\zpp=0.99$, $\zpm=-0.5$). The differences between the profiles for ($\zpp=0.89$, $\zpm=-0.54$) suggests that at least one of the two simulation protocols is not sampling equilibrium statistics.}
        \label{fig:sim_rdp_smallions_eqm1}
      \end{center}
    \end{figure}

\begin{figure}
      \begin{center}
        \includegraphics[width=0.8\textwidth]{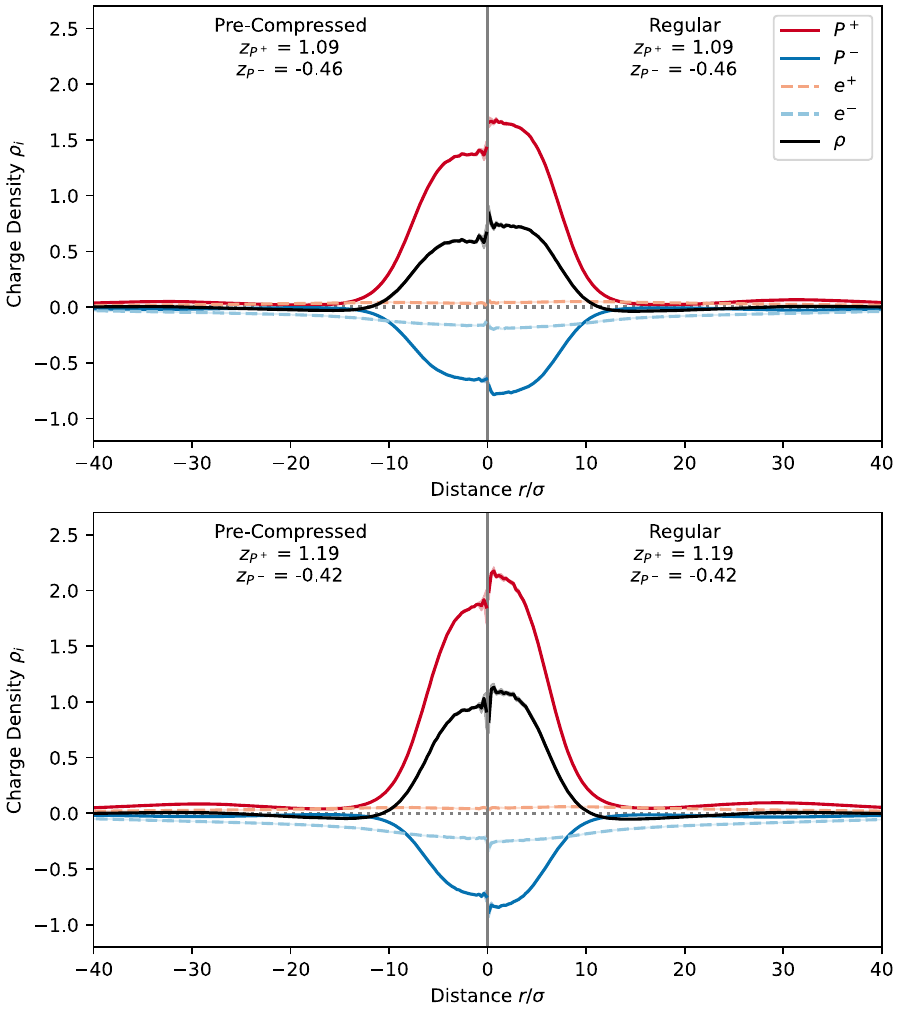}
        \caption{\textbf{Radial charge density profiles with reduced ion size for both pre-compressed and regular simulations.} Simulations with reduced ion size were performed using the pre-compressed (left) (subsection~\ref{sec:sim_equilibration}) and the regular (right) (subsection~\ref{sec:sim_procedure}) simulation procedures. Profiles are shown for the charge pairings of ($\zpp=1.09$, $\zpm=-0.46$) and ($\zpp=1.19$, $\zpm=-0.42$). The similarity in the profiles indicate that we are sampling equilibrium statistics.}
        \label{fig:sim_rdp_smallions_eqm2}
      \end{center}
    \end{figure}

  We also performed additional simulations to determine whether the simulations at a reduced ion size had reached equilibrium.
  To accomplish this, we ran the charge pairings simulated above---($\zpp=0.89$, $\zpm=-0.54$), ($\zpp=0.99$, $\zpm=-0.5$), ($\zpp=1.09$, $\zpm=-0.46$), ($\zpp=1.19$, $\zpm=-0.42$)---according to the procedure outlined in subsection~\ref{sec:sim_equilibration} (besides for replacing regular sized ion with these ion of a reduced size).
  As was done in subsection~\ref{sec:sim_equilibration}, we ran each system in triplicate and compared droplet size histograms (Figure~\ref{fig:sim_hist_smallions_eqm}) and radial charge density profiles (Figures~\ref{fig:sim_rdp_smallions_eqm1}, \ref{fig:sim_rdp_smallions_eqm2}) to the simulations performed with protocol from subsection~\ref{sec:sim_procedure} (but with reduced ion size).
  These comparisons demonstrate that we are likely sampling an equilibrated system for the ($\zpp=0.99$, $\zpm=-0.5$), ($\zpp=1.09$, $\zpm=-0.46$), and ($\zpp=1.19$, $\zpm=-0.42$) charge pairings.
  For the ($\zpp=0.89$, $\zpm=-0.54$) charge pairing, the pre-compressed simulations result in a single large sustained droplet for the entire duration of sampling whereas the simulations under the regular protocol produced a distribution of smaller droplets.
  This suggests that the results presented on the ($\zpp=0.89$, $\zpm=-0.54$) for simulations at a reduced ion size may not reflect thermodynamic equilibrium.
  
  \clearpage
  
  \subsection{Simulations in the overdamped limit}
  \label{sec:sim_overdamped}
  
\begin{figure}
      \begin{center}
        \includegraphics[width=0.8\textwidth]{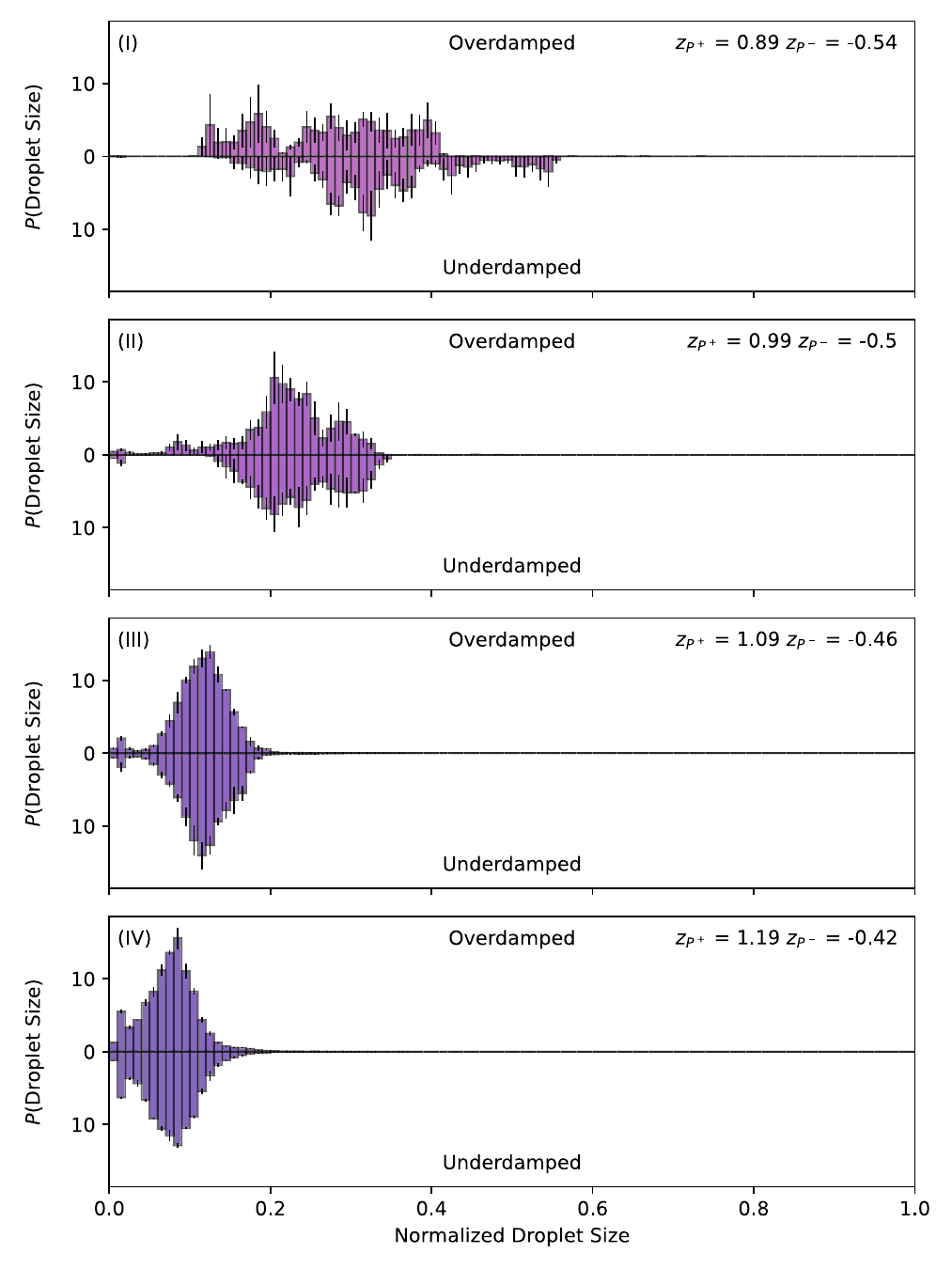}
        \caption{\textbf{Normalized droplet size distributions for overdamped versus underdamped simulations.} Histograms of droplet sizes are presented for simulations run in both the overdamped (top, damping of $\tau = 1.0$) and underdamped (bottom, damping of $\tau = 100.0$) limits for four charge asymmetries ($\zpp=0.89$, $\zpm=-0.54$), ($\zpp=0.99$, $\zpm=-0.5$), ($\zpp=1.09$, $\zpm=-0.46$), and ($\zpp=1.19$, $\zpm=-0.42$). The similarities between the droplet size distributions indicates that the results sampled from the simulation protocol described in subsection~\ref{sec:sim_procedure} are robust to the thermostat damping constant. The discrepancy in droplet sizes for the ($\zpp=0.89$, $\zpm=-0.54$) case may be due to finite sized effects which creates a broad distribution of sizes that are difficult to reproduce quantitatively between two sets of simulation triplicates.}
        \label{fig:sim_hist_langevin}
      \end{center}
    \end{figure}

\begin{figure}
      \begin{center}
        \includegraphics[width=0.8\textwidth]{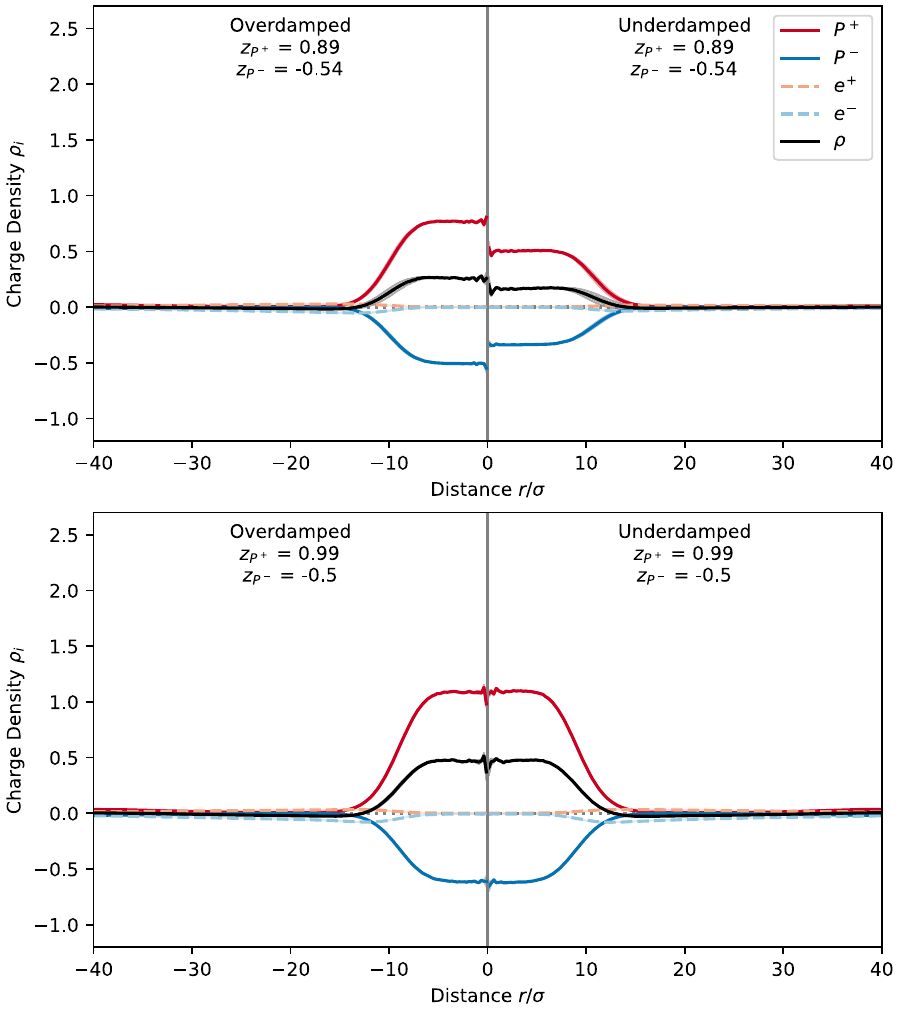}
        \caption{\textbf{Radial charge density profiles for overdamped versus underdamped simulations.} Radial charge density profiles are presented for simulations run in both the overdamped (top, damping of $\tau = 1.0$) and underdamped (bottom, damping of $\tau = 100.0$) limits for the charge asymmetries ($\zpp=0.89$, $\zpm=-0.54$) and ($\zpp=0.99$, $\zpm=-0.5$). Similarities in the profiles indicate that the results sampled from the simulation protocol described in subsection~\ref{sec:sim_procedure} are robust to thermostat damping constant. Discrepancies for the ($\zpp=0.89$, $\zpm=-0.54$) system may be attributed to finite sized effects which is also demonstrated by the broad distribution of droplet sizes in Figure~\ref{fig:sim_hist_langevin}.}
        \label{fig:sim_rdp_langevin1}
      \end{center}
    \end{figure}

\begin{figure}
      \begin{center}
        \includegraphics[width=0.8\textwidth]{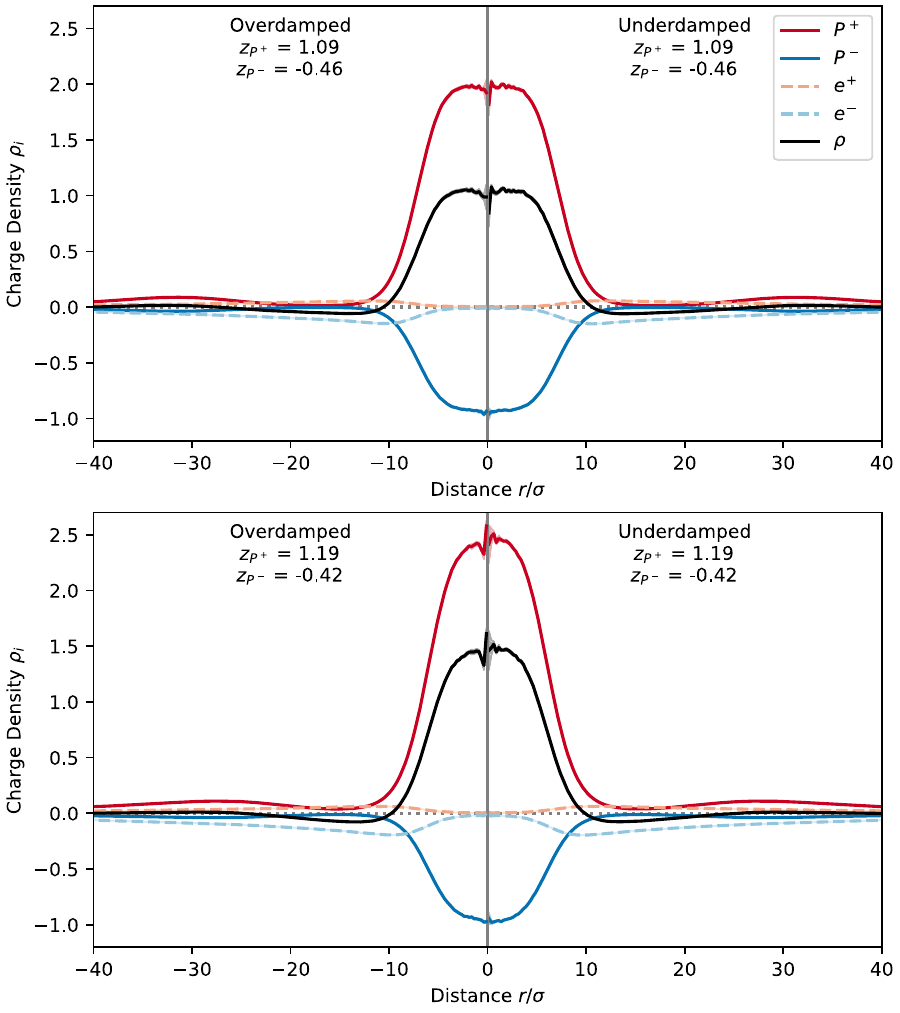}
        \caption{\textbf{Radial charge density profiles for overdamped versus underdamped simulations.} Radial charge density profiles are presented for simulations run in both the overdamped (top, damping of $\tau = 1.0$) and underdamped (bottom, damping of $\tau = 100.0$) limits for the charge asymmetries ($\zpp=1.09$, $\zpm=-0.46$) and ($\zpp=1.19$, $\zpm=-0.42$). Similarities in the profiles indicate that the results sampled from the simulation protocol described in subsection~\ref{sec:sim_procedure} are robust to thermostat damping constant.}
        \label{fig:sim_rdp_langevin2}
      \end{center}
    \end{figure}

  The Langevin thermostat used in the simulations discussed in the main text is underdamped(see the discussion of simulation protocol in subsection~\ref{sec:sim_procedure}).
  This should not change the results at equilibrium which can be demonstrated by simulating a subset of the simulations in the overdamped limit.
  The subset of simulations we choose to resimulate is ($\zpp=0.89$, $\zpm=-0.54$), ($\zpp=0.99$, $\zpm=-0.5$), ($\zpp=1.09$, $\zpm=-0.46$), and ($\zpp=1.19$, $\zpm=-0.42$).
  We run these simulations as discussed in subsection~\ref{sec:sim_procedure} with the exception of setting the Langevin thermostat damping constant to a damping of $1.0\tau$.
  Below we compare the droplet size histograms (Figure~\ref{fig:sim_hist_langevin}) and radial charge density profiles (Figures~\ref{fig:sim_rdp_langevin1}-\ref{fig:sim_rdp_langevin2}) between the sets of simulations at the two different damping constants.
  Near identical results are produced for the systems ($\zpp=0.99$, $\zpm=-0.5$), ($\zpp=1.09$, $\zpm=-0.46$), and ($\zpp=1.19$, $\zpm=-0.42$).
  In the system ($\zpp=0.89$, $\zpm=-0.54$) the overdamped simulations result in smaller droplets.
  This may be due to finite-sized effects at this charge pairing hampering the reproducibility of the distributions produced from the three simulation replicates.
  
  \clearpage

  \section{Details of the field theory}
  \setcounter{equation}{5}
  We here present mathematical and numerical details of the analysis of the field theory defined by Eqs. (1)--(3) in the main text.
  The main idea of the numerical minimization to obtain equilibrium states is to separate the entire system into $M$ different phases, which each can either be homogeneous or exhibit periodic patterns.
  Here, periodic patterns are described by the fields over one period together with the period length scale along each of the $d$ space dimensions.
  This is discussed in subsection~\ref{sec:FT_minimization}, which also includes the derivation that the interfacial energy is balanced by the electrostatic energy in equilibrium; see \Eqref{eqn:equivalence_electric_interface}.
  This approach naturally leads to a numerical minimization scheme, which we discuss in subsection~\ref{sec:FT_numerics}.
  We also discuss details of the linear stability analysis of the theory in subsection~\ref{sec:FT_linear_stability_analysis}, we quantify the distribution of ions inside and outside the droplets in subsection~\ref{sec:FT_ion_concentration}, and we provide additional figures in subsection~\ref{sec:FT_figures}.
  
  \subsection{Free energy minimization}
  \label{sec:FT_minimization}
  We consider a $d$-dimensional incompressible, isothermal mixture composed of $\Nc$ components with charge numbers $z_i$ for $i=1,...,\Nc$.
  The total free energy including local, interfacial, and long-ranged electrostatic interactions reads
  \begin{multline}
      F=\frac{\kbT}{v}\int_V\bigg[\sum_{i=1}^{\Nc}\frac{\phi_i(\br)}{l_i} \ln\phi_i(\br)+\frac12\sum_{i=1}^{\Nc}\sum_{j=1}^{\Nc}\chi_{ij}\phi_i(\br)\phi_j(\br) 
      + \frac{1}{2} \sum_{i=1}^{\Nc} \kappa_i|\nabla\phi_i(\br)|^2 \bigg] \mathrm{d}\vect r 
  \\
      +\int_V \biggl[-\frac{\varepsilon}{8\pi}|\nabla\psi(\br)|^2 + \frac{e\psi(\br)}{v}\sum_{i=1}^{\Nc} z_i\phi_i(\br)\biggr]\mathrm{d}\vect r
   \;.
  \end{multline}
  Using the Bjerrum length $\lb=e^2/(\varepsilon\kbT)$, $v\rightarrow v \lb ^3$, $V\rightarrow V \lb ^3$, $\kappa_i \rightarrow \kappa_i \lb^2$, $\nabla\rightarrow\nabla/\lb$, and $\psi \rightarrow \psi \kbT/e$, we  obtain the dimensionless free energy density $\hat{f}=Fv/(\kbT V)$,
  \begin{multline}
  \label{eq:hatf}
        \hat{f}=  \frac1{V}\int_V \bigg[\sum_i\frac{\phi_i(\br)}{l_i} \ln \phi_i(\br)+ \frac12\sum_{ij} \chi_{ij}\phi_i(\br)\phi_j(\br)+\frac12\sum_i{\kappa}_{i}  |\nabla \phi_i(\br)|^2
   -\frac{{v}}{8\pi}|\nabla\psi(\br)|^2+\psi(\br)\sum_iz_i\phi_i(\br)\bigg] \mathrm{d}\vect r
   \;.
  \end{multline}
  To study the coexistence of different phases, we consider a Gibbs ensemble of $M$ phases, where the fraction of volume of phase $\beta$ is $J_\beta$, implying $\sum_{\beta=1}^{M} J_\beta=1$.
  The average free energy density $\bar{f}= \sum_{\beta=1}^{M} J_{\beta} \hat{f}_\beta$ of the entire system then reads 
  \begin{multline}
      \bar f
       = \sum_{\beta=1}^{M}  \frac{J_{\beta}}{V_{\beta}}\int_{\Vbeta} \bigg[\sum_i\frac{\phi_i(\beta,\br)}{l_i} \ln \phi_i(\beta,\br)+ \frac12\sum_{ij} \chi_{ij}\phi_i(\beta,\br)\phi_j(\beta,\br)+\frac12\sum_i{\kappa}_{i}  |\nabla \phi_i(\beta,\br)|^2
       \notag \\
   -\frac{{v}}{8\pi}|\nabla\psi(\beta,\br)|^2+\psi(\beta,\br)\sum_iz_i\phi_i(\beta,\br)\bigg] \mathrm{d}\vect r
   \;.
  \end{multline}
  Additionally, there are several constraints, including mass conservation for each species,
  \begin{equation}
    \label{eqn:mass_conservation}
      \phibar_i=\sum_{\beta=1}^M  \frac{J_\beta}{\Vbeta}\int_\Vbeta \phi_i(\beta,\br)\mathrm{d}\br %
      \;,
  \end{equation}
  incompressibility everywhere in each phase, 
  \begin{equation}
      \sum_{i=1}^\Nc \phi_i(\beta,\br)=1
      \;,
  \end{equation}
  and charge neutrality in each phase,
  \begin{equation}
      \frac1 {\Vbeta} \int_{\Vbeta}\sum_{i=1}^\Nc z_i\phi_i(\beta,\br) \mathrm{d}\vect r=0
      \;.
  \end{equation}
  Using these constraints, the equilibrium coexisting states can be obtained by minimizing $\bar f$ over $\phi_i(\br)$, $\psi(\br)$, $J_\beta$, and the vector $\vect{L}_\beta$ describing the size of the compartment $\beta$ in all $d$ dimensions, which can be different in each phase.
  To alleviate the problem of negative volume fractions during the relaxation dynamics and to conserve the average volume fractions, we replace the logarithms associated with the translational entropies by introducing conjugated fields~$\omega_i(\beta, \br)$, akin to reference \cite{qiang2024scaling}.
  This leads to the extended average free energy density
  \begin{align}
  {f}&= \sum_{\beta=1}^{M}  \frac{J_{\beta}}{V_{\beta}}\int_{\Vbeta} \bigg[-\sum_i\phi_i(\beta,\br) \omega_i (\beta,\br)+ \frac12\sum_{ij} \chi_{ij}\phi_i(\beta,\br)\phi_j(\beta,\br)+\frac12\sum_i{\kappa}_{i}  |\nabla \phi_i(\beta,\br)|^2
  \notag \\
  &\quad-\frac{{v}}{8\pi}|\nabla\psi(\beta,\br)|^2+\psi(\beta,\br)\sum_iz_i\phi_i(\beta,\br)+\xi(\beta,\br)\biggl(\sum_j\phi_j(\beta,\br)-1\biggr)\bigg] \mathrm{d}\vect r 
  -\sum_i \frac{\phibar_i}{l_i}\ln Q_i
  \notag \\
  &\quad
  +\eta \biggl(\sum_\beta J_\beta -1\biggr) 
      + \sum_\beta J_\beta \zeta(\beta)\left(\frac1 {\Vbeta} \int_{\Vbeta}\sum_i z_i\phi_i(\beta,\br) \mathrm{d}\vect r\right) 
     + \sum_\beta J_\beta C\left(\frac1 {\Vbeta} \int_{\Vbeta}\sum_i z_i\phi_i(\beta,\br) \mathrm{d}\vect r\right)^2 ,
  \end{align}
  where
  \begin{eqnarray}
  \label{eq:Qi}
    Q_i=\sum_\beta J_\beta\frac1{\Vbeta}\int_{\Vbeta} e^{-\omega_i(\beta,\br) l_i}\mathrm{d}\vect r
   \;.
  \end{eqnarray}
  Here, $\omega_i(\beta,\br)$ are the conjugate fields of $\phi_i(\beta,\br)$, whereas $\xi(\beta,\br)$, $\eta$, and $\zeta(\beta)$ are the Lagrange multipliers for incompressibility, compartment volume conservation, and charge neutrality, respectively.
  The last term proportional to the constant $C$ is added to guide the convergence toward charge neutrality.
  This term does not contribute once the system is fully converged since global charge neutrality is fulfilled when \Eqref{eqn:mass_conservation} is obeyed. %

  The extremum of the free energy density $f$ with respect to $\omega_i(\beta,\br)$ provides
  \begin{equation}
    \frac{\delta {f}}{\delta\omega_i(\beta,\br) }=\Vbeta^{-1}\left(-\phi_i(\beta,\br){J_\beta}+{\phibar_i}\frac{e^{-\omega_i(\beta,\br) l_i}}{Q_i}{J_\beta}\right)=0
   \;,
  \end{equation}
  which gives
  \begin{equation}
    \label{eq:phi_i_r_beta}
    \phi_i(\beta,\br)={\phibar_i}\frac{e^{-\omega_i(\beta,\br) l_i}}{Q_i}
      \;.
  \end{equation}
  Note that this automatically satisfies material conservation,
  \begin{equation}
     \sum_{\beta=1}^{M} \frac{J_{\beta}}\Vbeta\int_\Vbeta\phi_i(\beta,\br) \mathrm{d}\vect r ={\phibar_i}
    \;.
  \end{equation}
  Inserting \Eqref{eq:phi_i_r_beta} together with the constraints into the free energy density $f$ gives exactly $\bar{f}$ except for a constant offset $-\sum_i\frac{\phibar_i}{l_i}\ln\phibar_i$, demonstrating the extremum of $f$ is equivalent to the extremum of $\bar{f}$, which we want to calculate.
  
  The extremum of $f$ with respect to  $\phi_i(\beta,\br)$ provides
   \begin{align}
   \label{eq:omega_i_r_beta}
    \omega_i(\beta,\br)=\sum_j \chi_{ij}\phi_j(\beta,\br)-{\kappa}_i\nabla^2\phi_i(\beta,\br)+\xi(\beta,\br)+z_i\psi(\beta,\br)+z_i\zeta(\beta)
   +2Cz_i\left(\frac1 {\Vbeta} \int_{\Vbeta}\sum_j z_j\phi_j(\beta,\br) \mathrm{d}\vect r\right) \;.
   \end{align}
  The extremum of $f$ with respect to  $\psi(\beta,\br)$ results in the Poisson's equation of electrostatics,
   \begin{equation}
   \label{eq:poisson}
    \nabla^2\psi(\beta,\br)=-\frac{4\pi}{{v}}\sum_iz_i\phi_i(\beta,\br)
   \;.
   \end{equation} 
  The extremum of $f$ with respect to $J_\beta$ provides
   \begin{align}
   \label{eq:eta}
    -\eta&=\frac1{\Vbeta}\int_0^\Vbeta  \bigg[-\sum_i\phi_i(\beta,\br) \omega_i (\beta,\br)+ \sum_{ij}\frac12 \chi_{ij}\phi_i(\beta,\br)\phi_j(\beta,\br)+\frac12\sum_i{\kappa}_{i}  |\nabla \phi_i(\beta,\br)|^2
    \notag \\
    &\quad-\frac{{v}}{8\pi}|\nabla\psi(\beta,\br)|^2+\psi(\beta,\br)\sum_iz_i\phi_i(\beta,\br)+\xi(\beta,\br)(\sum_j\phi_j(\beta,\br)-1)\bigg] \mathrm{d}\vect r 
    \notag \\
    &\quad-\sum_i \frac1{l_i} \frac1{\Vbeta}\int_\Vbeta \phi_i(\beta,\br)\mathrm{d}\vect r
  +\zeta(\beta)\frac1 {\Vbeta} \int_{\Vbeta}\sum_i z_i\phi_i(\beta,\br) \mathrm{d}\vect r
  +C\left(\frac1 {\Vbeta} \int_{\Vbeta}\sum_i z_i\phi_i(\beta,\br) \mathrm{d}\vect r\right)^2\;.
   \end{align}
  The extremum of $f$ with respect to $\zeta(\beta)$ gives charge neutrality in each compartment,
  \begin{equation}
  \label{eq:neutral}
    \frac1{\Vbeta} \int_{\Vbeta}\sum_i z_i\phi_i(\beta,\br) \mathrm{d}\vect r =0
   \;.
  \end{equation}
  The extremum of $f$ with respect to $\eta$ and $\xi(\beta,\br)$ simply gives incompressibility and volume conservation constraints,
   \begin{equation}
   \label{eq:incomp}
    \sum_i\phi_i(\beta,\br)=1\;,
   \end{equation}
   and 
   \begin{equation}
     \label{eq:jbeta}
      \sum_\beta {J_\beta}=1\;,
   \end{equation}
   respectively.
  The extremum of $f$ with respect to $\Lbeta^j$, the $j$-th component of $\vect L_\beta$, gives
  \begin{align}
    \frac{\mathrm{d} f^*}{\mathrm{d} \Lbeta^j}
   &=\left.\frac{\partial f}{\partial  \Lbeta^j}\right|_*
   +
   \int \sum_i \left.\frac{\delta f}{\delta \phi_i}\right\vert_*\frac{\partial \phi_i^*}{\partial\Lbeta^j}\mathrm{d}\br
    +\int \sum_i \left.\frac{\delta f}{\delta w_i}\right\vert_*\frac{\partial w_i^*}{\partial \Lbeta^j}\mathrm{d}\br
    +\int  \left.\frac{\delta f}{\delta\psi}\right\vert_*\frac{\partial \psi^*}{\partial \Lbeta^j}\mathrm{d}\br
    +\int  \left.\frac{\delta f}{\delta\xi}\right\vert_*\frac{\partial \xi^*}{\partial \Lbeta^j}\mathrm{d}\br+...
    \notag \\
    &=\left.\frac{\partial f}{\partial  \Lbeta^j}\right|_*
    \notag \\
    &=\frac{\partial}{\partial \Lbeta^j} \left[{J_\beta}\frac{1}{\Vbeta} \int_\Vbeta\biggl( \frac12\sum_i{\kappa}_{i}  |\nabla \phi_i(\beta,\br)|^2
    -\frac{{v}}{8\pi}|\nabla\psi(\beta,\br)|^2\biggr) \mathrm{d}\br\right]_*
   =0\;,
   \end{align}
   where the star denotes that quantities are evaluated for profiles that have been obtained by optimizing over all fields and parameters except $\Lbeta^j$.
   In particular, $f^*(\Lbeta^j)$ denotes the associated free energy density, which then only depends on $\Lbeta^j$.
   Using $f_\mathrm{int}^j=\Vbeta^{-1} \int_\Vbeta \frac12\sum_i{\kappa}_{i}  |\partial_{j} \phi_i(\beta,\br)|^2 \mathrm{d}\br$ and $f_\mathrm{\psi}^j=-\Vbeta^{-1} \int_\Vbeta \frac{{v}}{8\pi}(\partial_j\psi(\beta,\br))^2  \mathrm{d}\br$,
    we obtain 
   \begin{align}
     \label{eq:Lbeta}
     \frac{\mathrm{d} f^*}{\mathrm{d} \Lbeta^j}&=-J_\beta\frac{2}{\Lbeta^j}(f_\mathrm{int}^j+f_\mathrm{\psi}^j)=0\;,
    \end{align}
    which gives 
    \begin{eqnarray}
        f_\mathrm{int}^j=-f_\mathrm{\psi}^j\;.
    \end{eqnarray}
    Using Poisson's equation and integration by parts, we can express the electrostatic free energy density as
    \begin{eqnarray}
      \label{eqn:equivalence_electric_interface}
        f_{\mathrm{el}}=\frac{1}{\Vbeta} \int_\Vbeta \bigg[-\frac{{v}}{8\pi}(\nabla\psi(\beta,\br))^2  +\psi(\beta,\br) \sum_i z_i \phi_i(\beta,\br)\bigg] \mathrm{d}\br=-\sum_j f_\mathrm{\psi}^j,
    \end{eqnarray}
    and we thus find $f_{\mathrm{el}}=f_\mathrm{int}$ for all compartments, where $f_\mathrm{int}=\sum_j f_\mathrm{int}^j$ is the free energy density associated with interfaces.
    This indicates that the electrostatic energy is balanced by the interfacial energy in equilibrium.
   Specifically, in the 1D case, we have
   \begin{eqnarray}
   \label{eq:Lbeta_1D}
    \frac{\mathrm{d} f}{\mathrm{d} \Lbeta}&=&-{J_\beta}\frac{2}{\Lbeta} \frac1\Lbeta\int_0^\Lbeta\Bigg[ \frac12\sum_i{\kappa}_{i}  |\partial_x \phi_i(\beta,x)|^2
    -\frac{{v}}{8\pi}|\partial_x\psi(\beta,x)|^2\bigg] \mathrm{d}x 
    \nonumber\\
    &=&-{J_\beta}\frac{2}{\Lbeta} \frac1\Lbeta\int_0^\Lbeta\Bigg[ -\frac12\sum_i{\kappa}_{i}  \phi_i(\beta,x)\partial_x^2 \phi_i(\beta,x)
    +\frac{{v}}{8\pi}\psi(\beta,x)\partial_x^2\psi(\beta,x)\bigg] \mathrm{d}x =0
   \;.
  \end{eqnarray}
  In summary, we obtain the self-consistent equations \eqref{eq:Qi}, \eqref{eq:phi_i_r_beta}, 
  \eqref{eq:omega_i_r_beta}--\eqref{eq:jbeta},
  and \eqref{eq:Lbeta} to determine equilibrium states.

  \subsection{Numerical minimization method}
  \label{sec:FT_numerics}
  We designed an iterative scheme based on the self-consistent equations above, where we update all values of fields and variables based on their current approximated values, so the iteration converges to the free energy minimum.
  To improve convergence, we extensively use a simple mixing formula to improve numerical stability, i.e., for a generic variable $X$, we determine its value $X^{\text{new}}$ at the next iteration step as $X^\mathrm{new}=X^{\text{old}}+A_X(X^{\text{ideal}}-X^{\text{old}}) $, where $X^\mathrm{old}$ is the current value and $X^\mathrm{ideal}$ is the value suggested by the iteration scheme.
  Here, $A_X$ is an empirically determined mixing rate.
  
  In our numerical scheme, we first calculate $Q_i$ using \Eqref{eq:Qi}, and $\phi_i(\beta,\br)$ using \Eqref{eq:phi_i_r_beta}.
  We then calculate $\psi(\beta,\br)$ via
  \begin{equation}
        \label{eq:psi_i_algorithm}
    \psi(\beta,\br)=\mathcal F^{-1}\left[-\frac{1}{k^2}\mathcal F\left[-\frac{4\pi}{{v}}\sum_iz_i\phi_i(\beta,\br)\right]\right]\;,
  \end{equation}
  where $k$ is the wavenumber of the Fourier transform $\mathcal F$ and its inverse $\mathcal F^{-1}$.
  Next, we calculate $\xi(\beta,\br)$ using \Eqref{eq:omega_i_r_beta} and the incompressibility condition given by \Eqref{eq:incomp},
   \begin{align}
    \xi(\beta,\br)=-\frac1{\Nc}\sum_i\bigg[\sum_j \chi_{ij}\left(\phi_j(\beta,\br)+1-\sum_k\phi_k(\beta,\br)\right)-{\kappa}_i\nabla^2\phi_i(\beta,\br)
   +z_i\psi(\beta,\br)+z_i\zeta(\beta)
   \notag\\
   +2Cz_i\left(\frac1 {\Vbeta} \int_{\Vbeta}\sum_j z_j\phi_j(\beta,\br) \mathrm{d}\vect r\right)
   -\omega_i(\beta,\br)\bigg].
   \end{align}
   We now use \Eqref{eq:omega_i_r_beta} to update the ideal part of the new field, $\omega_i^\mathrm{ideal}(\beta,\br)$, via 
   \begin{eqnarray}
    \omega_i^\mathrm{ideal}(\beta,\br)=\sum_j \chi_{ij}\phi_j(\beta,\br)-{\kappa}_i\nabla^2\phi_i(\beta,\br)+z_i\psi(\beta,\br)+z_i\zeta(\beta)+\xi(\beta,\br)
   \;.
   \end{eqnarray}
  To ensure numerical stability, we update the new field $\omega^\mathrm{new}_i(\beta,\br)$ using a mixture of the old and the ideal one together with the kernel $R_i=\kappa_i$,
   \begin{eqnarray}
    \omega^\mathrm{new}_i(\beta,\br)=\omega_i(\beta,\br)+\mathcal F^{-1}\left[\frac1{A_\omega^{-1} + R_i k^2}\mathcal F\left[\omega^\mathrm{ideal}_i(\beta,\br)-{\omega}_i(\beta,\br)\right]\right]\;,
      \label{eq:omeganew}
   \end{eqnarray}
   where $A_\omega$ is an empirical mixing rate, typically around $0.001$. In fact, \Eqref{eq:omeganew} is 
  a simple mixing strategy together with a low pass filter, where the iteration is faster for low frequencies while it is slower slower for high frequencies to make the self-consistent iteration scheme more stable \cite{chen1998applications,thompson2004improved}.
  This approach empirically improved the convergence.
  We next use \Eqref{eq:eta} to update ${J_\beta}$. More specifically,
  using
  \begin{align}
        -\eta_\beta&=\frac1{\Vbeta}\int_0^\Vbeta  \bigg[-\sum_i\phi_i(\beta,\br) \omega_i (\beta,\br)+ \sum_{ij}\frac12 \chi_{ij}\phi_i(\beta,\br)\phi_j(\beta,\br)+\frac12\sum_i{\kappa}_{i}  |\nabla \phi_i(\beta,\br)|^2
    \notag \\
    &\quad-\frac{{v}}{8\pi}|\nabla\psi(\beta,\br)|^2+\psi(\beta,\br)\sum_iz_i\phi_i(\beta,\br)+\xi(\beta,\br)(\sum_j\phi_j(\beta,\br)-1)\bigg] \mathrm{d}\vect r 
    \notag \\
    &\quad-\sum_i \frac1{l_i} \frac1{\Vbeta}\int_\Vbeta \phi_i(\beta,\br)\mathrm{d}\vect r
  +\zeta(\beta)\frac1 {\Vbeta} \int_{\Vbeta}\sum_i z_i\phi_i(\beta,\br) \mathrm{d}\vect r
  +C\left(\frac1 {\Vbeta} \int_{\Vbeta}\sum_i z_i\phi_i(\beta,\br) \mathrm{d}\vect r\right)^2,
  \end{align}
  and 
  \begin{eqnarray}
    D_{\beta}=-\sum_{\gamma=1}^{M} \eta_{\gamma} J_{\gamma} + \eta_\beta\;,
  \end{eqnarray}
  we update $J_\beta$ via
  \begin{eqnarray}
    J^\mathrm{new}_\beta={J_\beta}+A_J D_{\beta}\;.
  \end{eqnarray}
  Here, $A_J$ is another empirical mixing rate, which is typically set to 0.001. Afterwards, we shift $J^\mathrm{new}_\beta$ to satisfy \Eqref{eq:jbeta} which reads $\sum_\beta J^\mathrm{new}_\beta=1$.
  Next, we update $\zeta(\beta)$ using \Eqref{eq:neutral},
  \begin{eqnarray}
    \zeta^\mathrm{new}(\beta)=\zeta(\beta)+A_{\zeta} {J_\beta}\frac1 {\Vbeta} \int_{\Vbeta}\sum_i z_i\phi_i(\beta,\br) \mathrm{d}\br \;,
  \end{eqnarray}
  where $A_{\zeta}$ is the third mixing rate, which is set to $10$.
  Finally, in 1D, we update $L_\beta$ using \Eqref{eq:Lbeta_1D}, via
  \begin{eqnarray}
    L^\mathrm{new}_\beta&=&L_\beta + A_L {J_\beta}\frac{2}{\Lbeta} \frac1\Lbeta\int_0^\Lbeta\bigg[ -\frac12\sum_i{\kappa}_{i}  \phi_i(\beta,x)\partial_x^2 \phi_i(\beta,x)
    +\frac{{v}}{8\pi}\psi(\beta,x)\partial_x^2\psi(\beta,x)\bigg] \mathrm{d}x \;,
  \end{eqnarray}
  where $A_L $ is an empirical mixing rate around $10$.
  We use our iterative scheme to find equilibrium states where each phase is described by fields  $\phi_i(\br)$ and $\psi(\br)$ together with optimized sizes $L_\beta$. 
  To produce the results of the main text, we use $\Nc=5$, assume $M=2$, and discretize fields at $128$ points along the single dimension.
  We have checked the correctness at selected parameter values using $M=8$ phases; All convergent results are consistent with those using $M=2$.
  
  \subsection{Linear stability analysis}
  \label{sec:FT_linear_stability_analysis}
\begin{figure}
      \begin{center}
        \includegraphics[width=0.5\textwidth]{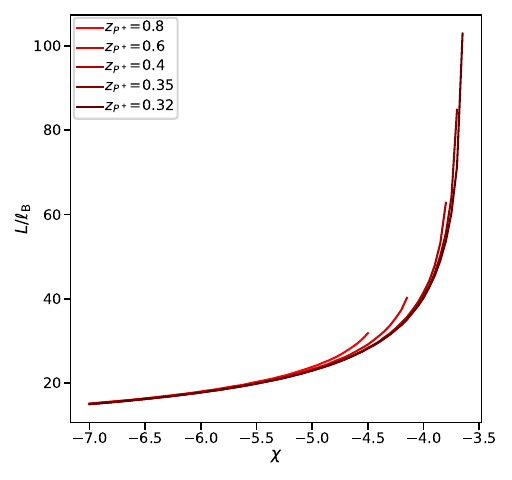}
        \caption{ 
        The most unstable length scale $L$ predicted from linear stability analysis as a function of the interaction parameter~$\chi$.
        Parameters are the same as in Fig. 4 of the main text.
        }
        \label{fig:SI_spinodal_L_chi}
      \end{center}
    \end{figure}

  We here present the linear stability analysis of the field theory around the uniform state $\phi_i(x)=\phibar_i$.
  We perturb $\phibar_i$ by a small value, setting $\phi_i(x)=\phibar+\epsilon_i \cos(qx)$, where $\epsilon_i$ is the perturbation amplitude and $q$ is the associated wave number. By evaluating \Eqref{eq:hatf}, taking the second-order derivative of $\hat{f}$ with respect to $\epsilon_i$, and then taking the limit $\epsilon_i\rightarrow 0$, we obtain the $4\times 4$ Hessian matrix
  \begin{equation}
      H_{ij}=q^2\left(\frac{1}{\phibar_il_i}\delta_{ij}+\frac{1}{\bar{\phi}_\mathrm{S}}\right)+q^4\kappa_i\delta_{ij}+\frac{4\pi}{v}z_iz_j
      \;.
  \end{equation}
  For the homogeneous state to be stable, all eigenvalues $\lambda_i$ of the Hessian matrix must be positive for all $q$, which we check numerically.
  If the homogeneous state is unstable, we numerically calculate the eigenvalues as a function of $q$ and determine the minimal value at $q=q_\mathrm{min}$.
  The length scale associated with this most unstable mode is $2\pi/q_\mathrm{min}$ and shown in Fig. 4 of the main text.
  It provides an estimate for the length scale originating when an unstable homogeneous system is prepared, and it sometimes provides a reasonable estimate for the final pattern length scale.
  To test this, we show the most unstable length scale for small charge asymmetry in \figref{fig:SI_spinodal_L_chi}, indicating that this length strongly depends on $\chi$, while the charge asymmetry has a weak effect, in contrast to the results for the equilibrium length scale presented in the main text.

  \subsection{Quantification of ion concentration}
  \label{sec:FT_ion_concentration}
  
\begin{figure}
      \begin{center}
        \includegraphics[width=0.5\textwidth]{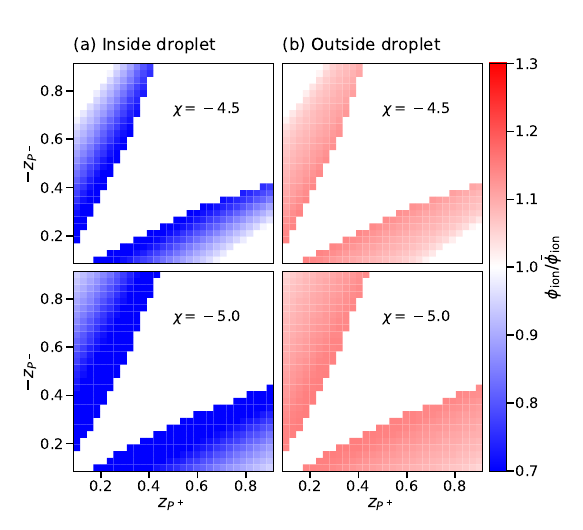}
        \caption{ 
        The relative mean volume fraction of ions, $\phi_\mathrm{ion}/\bar{\phi}_\mathrm{ion}$, for (a) inside and (b) outside the droplet as a function of the charge numbers of the polymers. Here, $\phi_\mathrm{ion}$ is the mean volume fraction of all small ions ($e^+$ and $e^-$) inside or outside the droplet, and $\bar{\phi}_\mathrm{ion}=\bar{\phi}_{e^+}+\bar{\phi}_{e^-}$.
        The inside of the droplet is defined as the region where \(\phiS < \frac12(\phiS^\mathrm{min} + \phiS^\mathrm{max}) \), with \(\phiS^\mathrm{min}\) and \(\phiS^\mathrm{max}\) representing the minimum and maximum values of the solvent volume fraction, respectively.
        Conversely, regions where \(\phiS \geq \frac12(\phiS^\mathrm{min} + \phiS^\mathrm{max}) \) are considered outside the droplet.
        }
        \label{fig:SI_ions_zA_zC}
      \end{center}
    \end{figure}

\begin{figure}
      \begin{center}
        \includegraphics[width=0.5\textwidth]{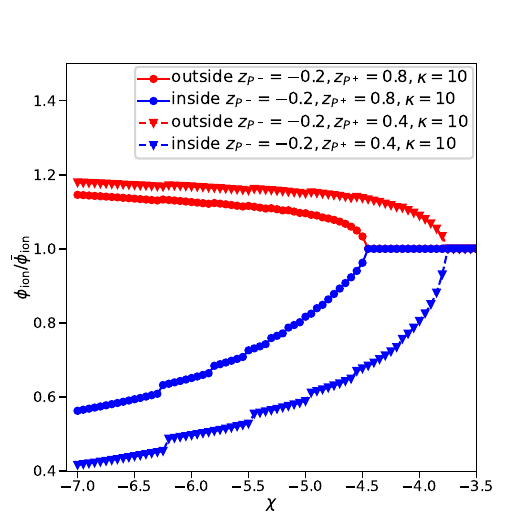}
        \caption{ 
        The relative mean volume fraction of ions, $\phi_\mathrm{ion}/\bar{\phi}_\mathrm{ion}$, as a function of the interaction strength~$\chi$ for various charge numbers of the polymers.
        A stronger short-range attraction (more negative $\chi$) results in a greater expulsion of ions from the droplet.
        }
        \label{fig:SI_ions_chi}
      \end{center}
    \end{figure}

  One of the main messages of the main text is that ions are unevenly distributed in the patterned phase, with a higher concentration of ions outside the droplet compared to the interior.
  We here quantify this effect by calculating the average fraction of ions inside droplets and outside droplets separately.
  \figref{fig:SI_ions_zA_zC} shows that ions are enriched outside droplets (right column) while they are depleted inside droplets (left columns) in the patterned phase.
  This disparity arises because the short-range attraction between polymers effectively expels ions from the droplet.
  The strength of this expulsion increases with stronger polymer attractions, (larger $-\chi$) as illustrated in \figref{fig:SI_ions_chi}.
  However, the expulsion weakens as charge asymmetry increases; see \figref{fig:SI_ions_zA_zC}.


  \subsection{Additional result figures}
  \label{sec:FT_figures}
    
\begin{figure*}
      \centering
      \includegraphics[width=1\textwidth]{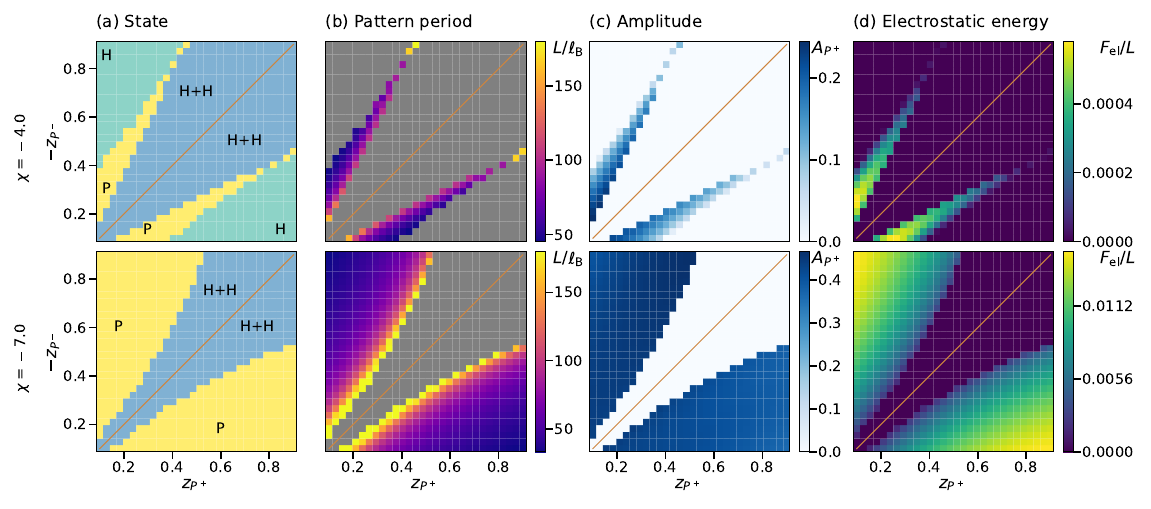}
      \caption{
      Results analogous to Fig. 3 in the main text for $\chi=-4$ (upper row) and $\chi=-7$ (lower row).
      (a) Phase diagram as a function of the charge numbers $\zC$ and $\zA$ of the polymers revealing parameter regions with the coexistence of two homogeneous phases (H+H), patterned phases (P), a single homogeneous phase (H).
      (b) Period~$L$ in the patterned phase corresponding to panel a.
      (c) Amplitude $A_{P^+} = \max(\phiC) - \min(\phiC)$ corresponding to panel a.
      (d) Electrostatic energy $F_\mathrm{el}$ corresponding to panel a.
      }
      \label{fig:types_MF_SI} 
    \end{figure*}

\begin{figure}
      \begin{center}
        \includegraphics[width=0.95\textwidth]{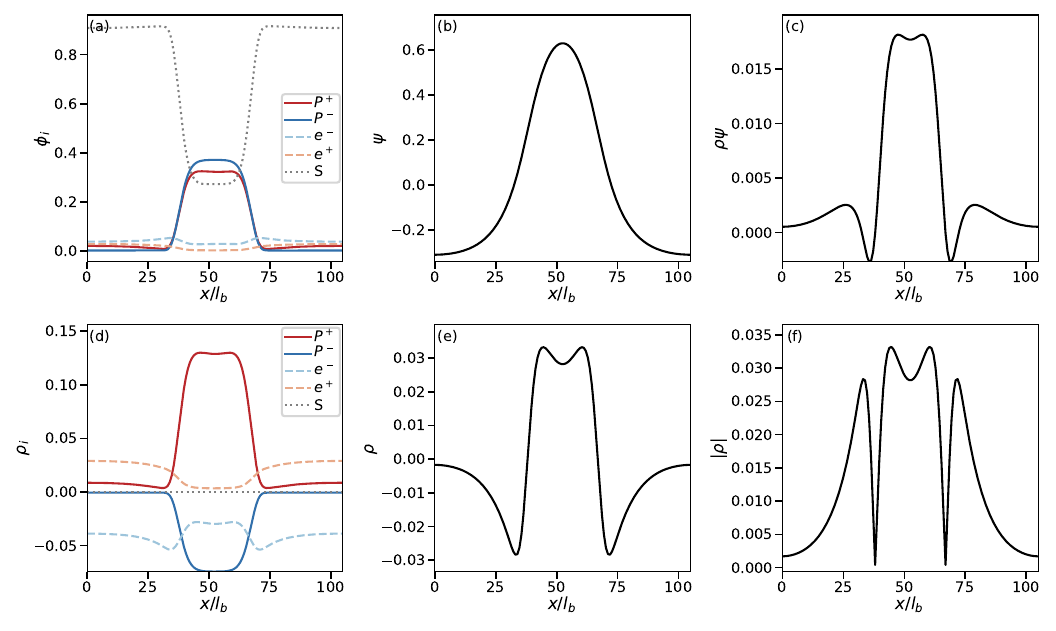}
        \caption{
        Additional details for the state for weak charge asymmetry shown in Fig. 2a in the main text.
        Parameters are $\zC=0.4$, $\zA=-0.2$, $\chi=-5$, and $\kappa=10$.
        (a) Volume fractions $\phi_i$ as a function of position $x$.
        (b) Electrostatic potential $\psi(x)$.
        (c) Local electrostatic energy density $\rho\psi(x)$.
        (d) Charge density $\rho_i(x)=z_i\phi_i(x)$ for each species.
        (e) Total charge density $\rho(x)=\sum_{i}\rho_i(x)$.
        (f) Absolute value of total charge density $|\rho(x)|$.
        }
        \label{fig:profile1_MF_SI}
      \end{center}
    \end{figure}

\begin{figure}
      \begin{center}
        \includegraphics[width=0.95\textwidth]{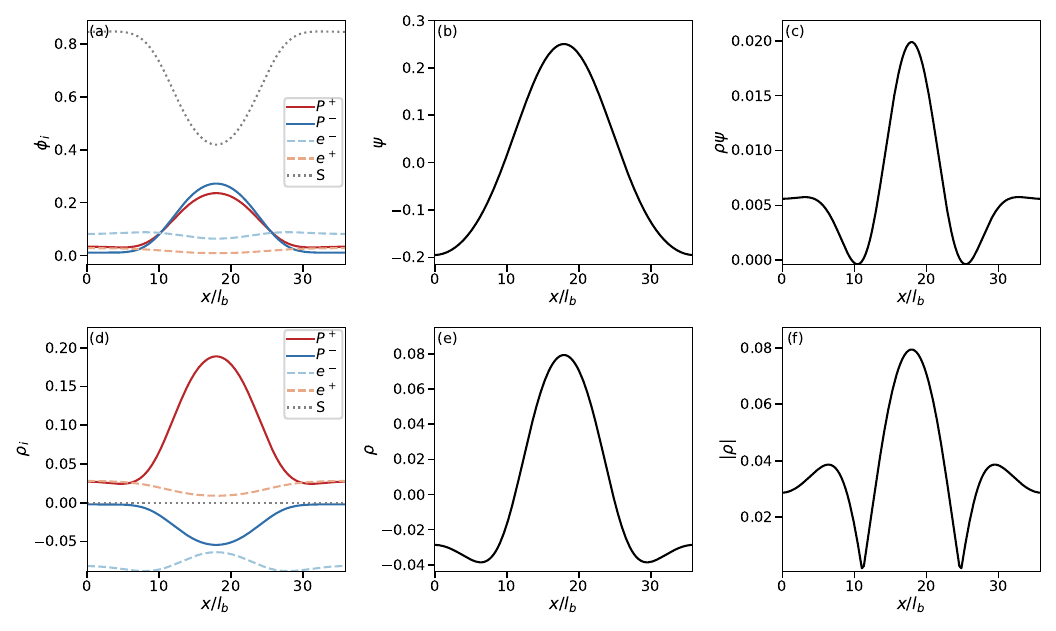}
        \caption{%
        Additional details for the state for weak charge asymmetry shown in Fig. 2b in the main text.
        Parameters are $\zC=0.8$, $\zA=-0.2$, $\chi=-5$, and $\kappa=10$.
        (a) Volume fractions $\phi_i$ as a function of position $x$.
        (b) Electrostatic potential $\psi(x)$.
        (c) Local electrostatic energy density $\rho\psi(x)$.
        (d) Charge density $\rho_i(x)=z_i\phi_i(x)$ for each species.
        (e) Total charge density $\rho(x)=\sum_{i}\rho_i(x)$.
        (f) Absolute value of total charge density $|\rho(x)|$.
        }
        \label{fig:profile2_MF_SI}
      \end{center}
    \end{figure}

\begin{figure}
      \begin{center}
        \includegraphics[width=0.95\textwidth]{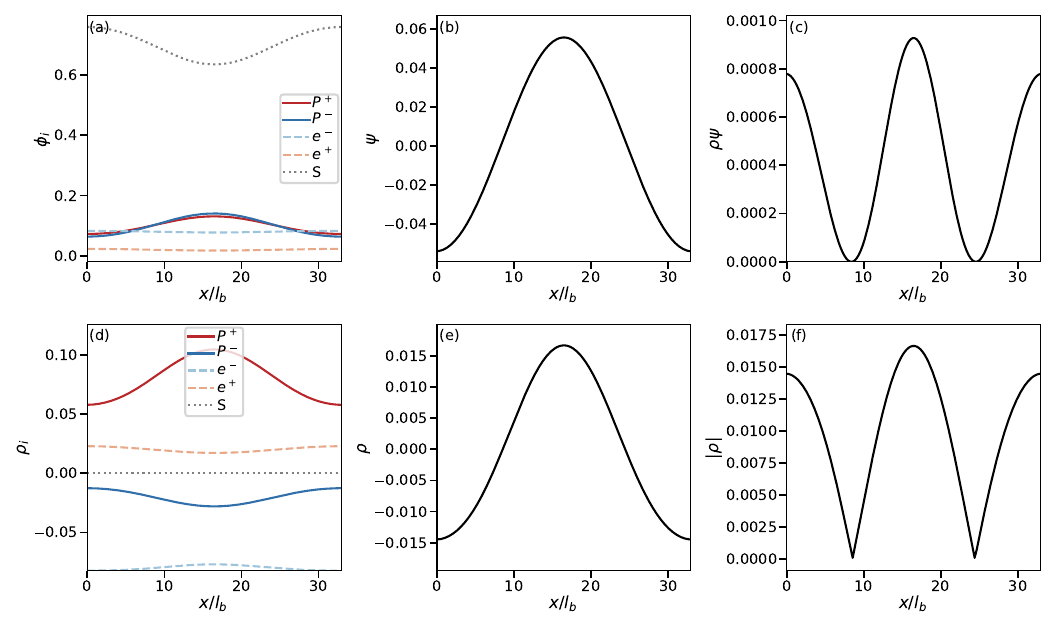}
        \caption{%
        Typical profile close to continuous transition.
        Parameters are $\zC=0.8$, $\zA=-0.2$, $\chi=-4.5$ (the critical point is at $\chi^*\approx-4.46$), and $\kappa=10$.
        (a) Volume fractions $\phi_i$ as a function of position $x$.
        (b) Electrostatic potential $\psi(x)$.
        (c) Local electrostatic energy density $\rho\psi(x)$.
        (d) Charge density $\rho_i(x)=z_i\phi_i(x)$ for each species.
        (e) Total charge density $\rho(x)=\sum_{i}\rho_i(x)$.
        (f) Absolute value of total charge density $|\rho(x)|$.
        }
        \label{fig:profile3_MF_SI}
      \end{center}
    \end{figure}

\begin{figure}
      \begin{center}
        \includegraphics[width=0.9\textwidth]{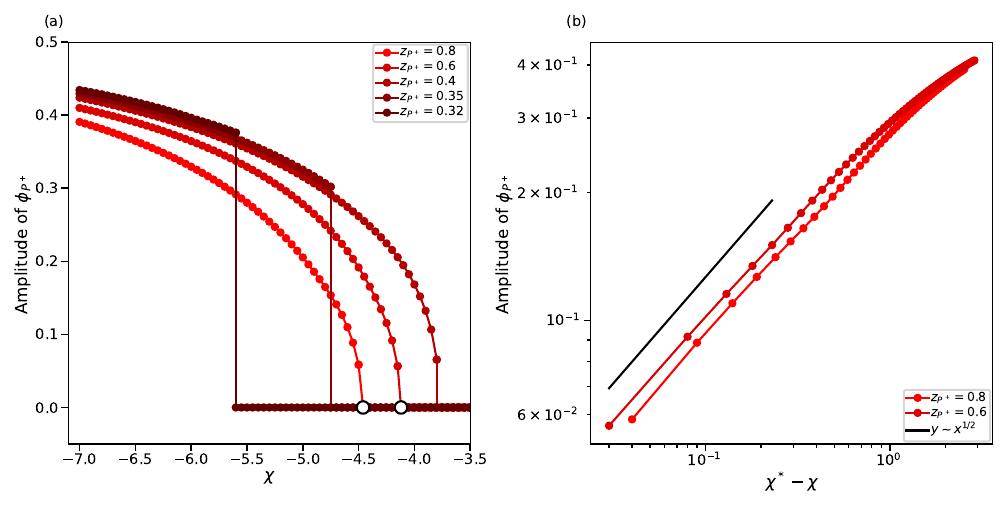}
        \caption{\textbf{Amplitude decreases as $\chi$ increases.} (a) The amplitude of fraction $\phi_{P^+}$ of the positively-charged polymer decreases as the interaction parameter $\chi$ increases. The circle indicates the critical point at $\chi=\chi^*$ where $\chi^*\approx -4.46$ for $\zC=0.8$ and $\chi^*\approx -4.12$ for $\zC=0.6$.
        (b) Rescaled data of panel a, now shown as a function of the relative interaction strength $\chi^*-\chi$. The exponent $\frac12$ is consistent with the critical exponent in mean-field theory, indicating that the transition is continuous.
        Parameters are $\zA=-0.2$ and $\kappa=10$. Remaining parameters are the same as in Fig. 2 in the main text.}
        \label{fig:amplitude_MF_SI}
      \end{center}
    \end{figure}

\begin{figure}
      \begin{center}
        \includegraphics[width=0.9\textwidth]{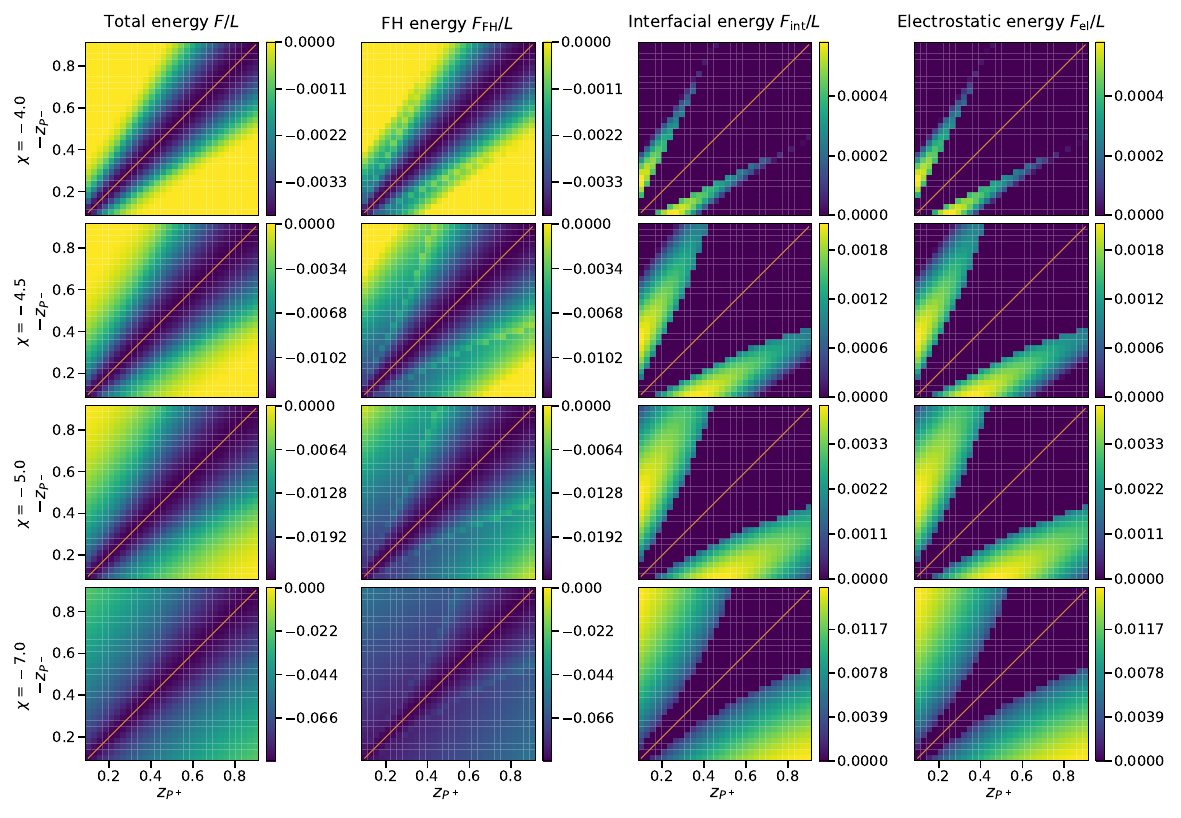}
        \caption{\textbf{Different contributions to the free energy.}
        Total energy density $F$, and its three contributions $F_\mathrm{FH}$, $F_\mathrm{int}$, and $F_\mathrm{el}$, normalized to the size $L$ as a function of the charge numbers $\zC$ and $\zA$ of the polymers for various interaction strengths~$\chi$ (across rows).
        To highlight the details, $F$ and $F_\mathrm{FH}$ have been shifted by the free energy $F(\phi_i(x)=\bar\phi_i$) of the respective homogeneous states.
        Consistent with \Eqref{eqn:equivalence_electric_interface}, $F_\mathrm{int}$ and $F_\mathrm{el}$ are always identical. 
        }
        \label{fig:energy_SI}
      \end{center}
    \end{figure}

  Figure \ref{fig:types_MF_SI} shows phase diagram of states, pattern periods, amplitudes, and electrostatic energy  for $\chi=-4$ and $\chi=-7$, analogous to Fig. 3(b) in the main text. Figures \ref{fig:profile1_MF_SI} and \ref{fig:profile2_MF_SI} provide additional details for states depicted in Fig. 2 of the main text.
  Figure \ref{fig:profile3_MF_SI} shows a typical profile near the continuous transition.
  The nature of this transition can be investigate from the dependence of the amplitude of the volume fraction of the negatively-charged polymer~$P_-$ as a function of $\chi$; see \figref{fig:amplitude_MF_SI}.
  The different contributions to the free energy are shown in \figref{fig:energy_SI}.  The last two columns, in particular, demonstrate that the interfacial energy $F_\mathrm{int}$ is equal to the electrostatic energy $F_\mathrm{el}$.
  
\begin{figure*}
      \centering
      \includegraphics[width=1\textwidth]{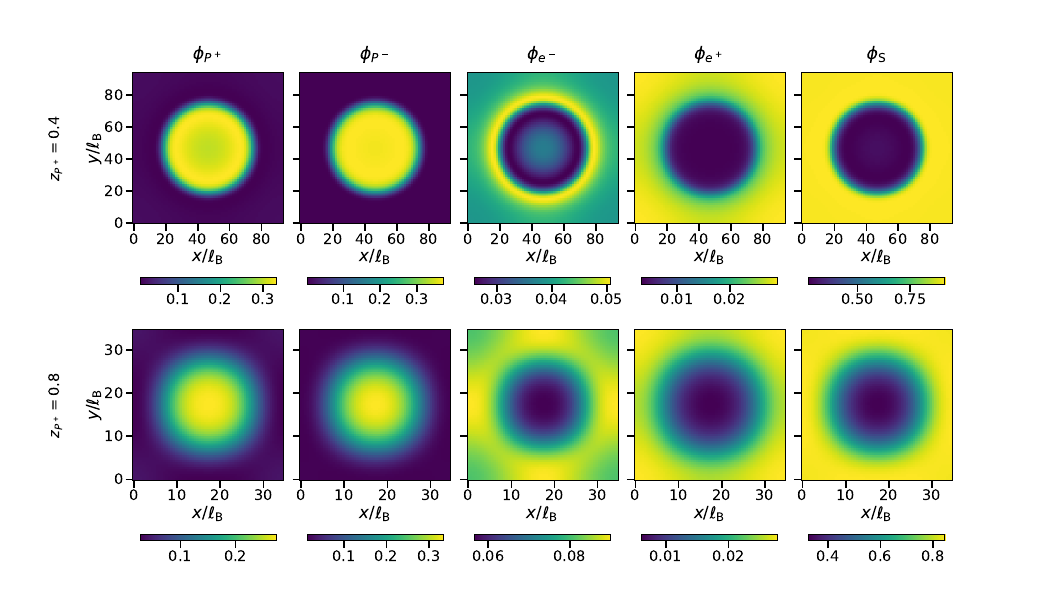}
      \caption{
      Two-dimensional profiles the volume fractions $\phi_i(x,y)$ of all species (across columns) predicted by the field theory for two different charge numbers~$z_{P^+}$ of the positively-charged polymers (across rows). The remaining parameters are the same as in Fig. 2 of the main text.
      }
      \label{fig:2d_phis} 
    \end{figure*}

\begin{figure*}
      \centering
      \includegraphics[width=1\textwidth]{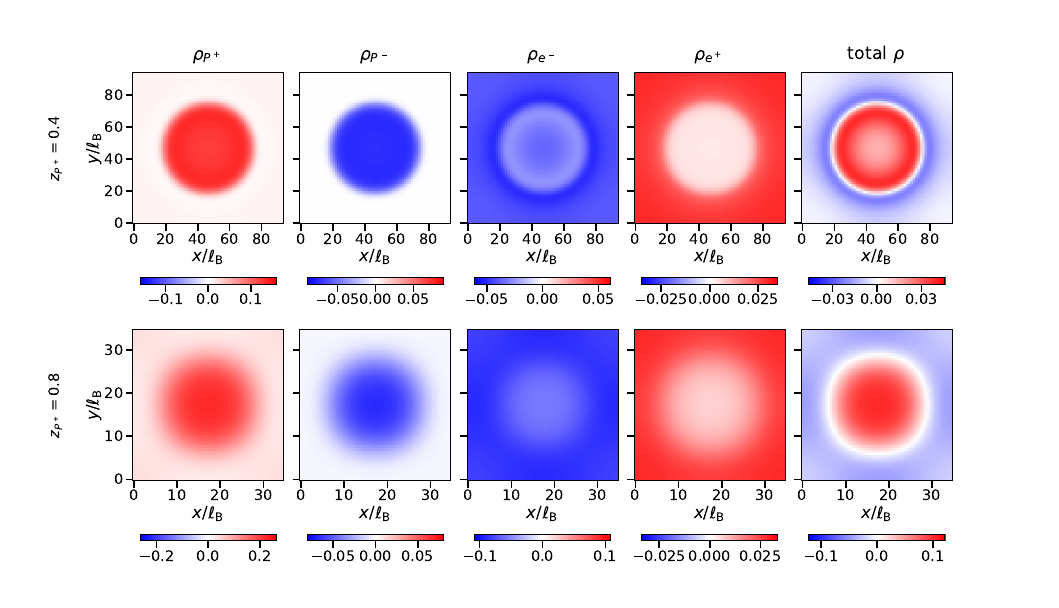}
      \caption{
      Two-dimensional profiles of charge densities, $\rho_i(x,y)=z_i\phi_i(x,y)$, of charged species (across columns) predicted by the field theory for two charge numbers~$z_{P^+}$ of the positively-charged polymers (across rows). Last column shows the total charge density $\sum_i \rho_i(x,y)$. 
      }
      \label{fig:2d_rhos} 
    \end{figure*}

  To check whether dimensionality matters, we also obtain the patterned phase by solving the two-dimensional mean-field theory equations.
  We show the profiles of volume fractions in \figref{fig:2d_phis} and charge densities in \figref{fig:2d_rhos}, corresponding to the same states as in Fig. 2 of the main text. The 2D disk patterns qualitatively agree with the 1D results; Notably, we also observe a dip inside the droplet at $z_{P^+}=0.4$.
  Given the low total volume fractions of polymers, $\phiC+\phiA=0.2$, we expect spheres, rather than lamellar or cylindrical structures, will form in 3D within our parameter regime.


\end{document}